\documentstyle[preprint,aps,prb]{revtex} 

\begin{document} 
\title{Hydrodynamics for a granular binary mixture at low density} 
\author{Vicente Garz\'{o}\footnote[1]{Electronic mail: vicenteg@unex.es}} 
\address{Departamento de F\'{\i}sica, Universidad de Extremadura, E-06071 \\ 
Badajoz, Spain} 
\author{James W. Dufty\footnote[2]{Electronic mail: dufty@phys.ufl.edu}} 
\address{Department of Physics, University of Florida, Gainesville, \\ 
Florida 32611} 
\date{\today} 
\maketitle 
\tighten
\begin{abstract} 
Hydrodynamic equations for a binary mixture of inelastic hard spheres are 
derived from the Boltzmann kinetic theory. A normal solution is 
obtained via the Chapman-Enskog method for states near the local homogeneous 
cooling state. The mass, heat, and momentum fluxes are determined to first 
order in the spatial gradients of the hydrodynamic fields, and the 
associated transport coefficients are identified. In the same way as for 
binary mixtures with elastic collisions, these coefficients are determined 
from a set of coupled linear integral equations. Practical evaluation is 
possible using a Sonine polynomial approximation, and is illustrated here by 
explicit calculation of the relevant transport coefficients:  the mutual 
diffusion, the pressure diffusion, the thermal diffusion, the shear viscosity, the 
Dufour coefficient,  the thermal conductivity, and the pressure energy coefficient. 
All these coefficients are given in terms of  the restitution coefficients 
and the ratios of mass, concentration, and particle sizes. Interesting and 
new effects arise from the fact that the reference states for the two 
components have different partial temperatures, leading to additional 
dependencies of the transport coefficients on the concentration. The results 
hold for arbitrary degree of inelasticity and are not limited to specific 
values of the parameters of the mixture. Applications of this theory will be 
discussed in subsequent papers.  
\end{abstract} 
 
\draft 
\pacs{PACS number(s): 45.70.Mg, 05.20.Dd, 51.10.+y} 
 
\bigskip \narrowtext 
%\newpage 
 
\section{Introduction} 
 
\label{sec1} 
 
The qualitative properties of a granular gas whose dynamics is dominated by 
pair-wise collisions between the grains can be described by a Boltzmann 
equation, suitably modified to describe the more complex two particle 
collisions. In the simplest model the grains are taken to be smooth, hard 
spheres with inelastic collisions. In recent years, the derivation of 
hydrodynamic equations for a one component granular gas from this idealized 
Boltzmann description has been worked out in detail to Navier-Stokes order,  
\cite{BDKS98} with explicit expressions for the transport coefficients as 
a function of the degree of dissipation (the restitution coefficient). These 
recent results improve upon earlier studies \cite{JR85,LSJCh84,GS95} by 
providing expressions that are accurate even for strong dissipation. As a 
consequence, there are now precise predictions from the Boltzmann equation 
suitable for detailed comparison with Monte Carlo simulation, molecular 
dynamics simulation, and the evolving new class of controlled experiments. 
This analysis for the one component system also has been extended to dense 
gases described by the Enskog equation. \cite{GD199} Similar studies for 
multicomponent granular gases are more scarce. Existing work on 
multicomponent transport appears to be based on weak dissipation 
approximations. \cite{JM89,Z95,AW98,WA99} Our objective is to 
provide a description of hydrodynamics in binary granular mixtures with a 
comparable accuracy to that for the one component system, valid over the 
broadest parameter range including strong dissipation. The reference 
homogeneous cooling state for a binary mixture has been discussed in detail 
recently \cite{GD299} to provide the proper basis for analysis of transport 
due to spatial inhomogeneities. Those results are used here to describe 
mass, heat, and momentum transport. The expressions for the distribution 
functions, fluxes, and transport coefficients are exact to Navier-Stokes 
order (within the context of the Boltzmann equation). 
 
The hydrodynamic equations for a binary mixture at low density are derived 
from the coupled set of Boltzmann equations for the two species in the same 
manner as for a one component system. The solutions for the distribution 
functions are expanded about a local homogeneous cooling state that is 
analogous to the local equilibrium state for a gas with elastic collisions. 
The expansion is in powers of the spatial gradients of the hydrodynamic 
fields (e.g., species densities, temperature, and flow velocity) and is an 
extension of the familiar Chapman-Enskog procedure for elastic collisions. 
The primary technical complication for inelastic collisions is an inherent 
time dependence of the reference state due to collisional cooling. In a one 
component system this occurs through the time dependence of the temperature 
defined in terms of the mean square velocity for the homogeneous cooling 
distribution. For a two component system the temperature is defined in terms 
of the algebraic average of the mean square velocities for the two 
distributions. In the case of elastic collisions the average temperature is 
the same as the kinetic temperatures for each species in the local 
equilibrium state. However, a surprising result of the study in Ref.\ \onlinecite {GD299} is that these temperatures are all different for 
inelastic collisions. This does not mean that there are additional 
hydrodynamic degrees of freedom since their cooling rates are the same and 
consequently, the partial kinetic temperatures still can be expressed in 
terms of the average temperature. But the relationships between these 
temperatures are functions of the densities for the two species and lead to 
a new dependence of the reference cooling states on these hydrodynamic 
variables. The consequences of this effect for the transport coefficients is 
significant, as shown below. 
 
The hydrodynamic equations for a binary mixture are somewhat more 
complicated than for the one component case: six coupled equations with 
eight transport coefficients. The irreversible (dissipative) parts of the 
mass, heat, and momentum fluxes are calculated to leading order in the 
spatial gradients of the hydrodynamic fields. For systems with elastic 
collisions the specific set of gradients contributing to each flux is 
restricted by fluid symmetry, time reversal invariance (Onsager relations), 
and the form of the entropy production.\cite{GM84} For inelastic collisions only fluid 
symmetry applies so there is greater freedom in representing the fluxes and 
identifying associated transport coefficients. This is discussed further in 
Section \ref{sec3} where the independent gradients are chosen to be those 
for the concentration of species one, the pressure, the temperature and the 
components of the flow field, with eight independent scalar transport 
coefficients. Using the Chapman-Enskog expansion the solutions to the 
Boltzmann equations are obtained to leading order in these gradients, and 
the transport coefficients are expressed in terms of the solutions to a set 
of coupled linear integral equations. 
 
The plan of the paper is as follows. In Sec.\ \ref{sec2}, the coupled set of 
Boltzmann equations and the corresponding hydrodynamic equations are 
recalled. The Chapman-Enskog expansion adapted to the inelastic binary 
mixtures is described in Sec.\ \ref{sec3} to construct the distribution 
function to linear order in the gradients. This solution is used to 
calculate the fluxes and identify associated transport coefficients. A 
Sonine polynomial approximation is applied to solve the linear integral 
equations defining selected transport coefficients in Sec.\ \ref{sec4}. 
We get explicit expressions for these transport coefficients in terms of the 
restitution coefficients and the masses, concentrations, and sizes of the 
constituents of the mixture.  
Finally, the results are summarized and discussed in Sec.\ \ref 
{sec5}.

\section{Boltzmann equation and conservation laws} 
 \label{sec2} 
 
Consider a binary mixture of smooth hard spheres of masses $m_{1}$ and $% 
m_{2} $, and diameters $\sigma _{1}$ and $\sigma _{2}$. The inelasticity of 
collisions among all pairs is characterized by three independent constant 
coefficients of normal restitution $\alpha _{11}$, $\alpha _{22}$, and $% 
\alpha _{12}=\alpha _{21}$, where $\alpha _{ij}$ is the restitution 
coefficient for collisions between particles of species $i$ and $j$. In the 
low-density regime, the distribution functions $f_{i}({\bf r},{\bf v};t)$ $% 
(i=1,2)$ for the two species are determined from the set of nonlinear 
Boltzmann equations \cite{BDS97}  
\begin{equation} 
\left( \partial _{t}+{\bf v}_{1}\cdot \nabla \right) f_{i}({\bf r},{\bf v} 
_{1},t)=\sum_{j}J_{ij}\left[ {\bf v}_{1}|f_{i}(t),f_{j}(t)\right] \;. 
\label{2.1} 
\end{equation} 
The Boltzmann collision operator $J_{ij}\left[ {\bf v}_{1}|f_{i},f_{j}\right] 
$ describing the scattering of pairs of particles is  
\begin{eqnarray} 
J_{ij}\left[ {\bf v}_{1}|f_{i},f_{j}\right] &=&\sigma _{ij}^{2}\int d{\bf v} 
_{2}\int d\widehat{\bbox {\sigma }}\,\Theta (\widehat{{\bbox {\sigma }}} 
\cdot {\bf g}_{12})(\widehat{\bbox {\sigma }}\cdot {\bf g}_{12})  \nonumber 
\\ 
&&\times \left[ \alpha _{ij}^{-2}f_{i}({\bf r},{\bf v}_{1}^{\prime 
},t)f_{j}( {\bf r},{\bf v}_{2}^{\prime },t)-f_{i}({\bf r},{\bf v}% 
_{1},t)f_{j}({\bf r}, {\bf v}_{2},t)\right] \;,  \label{2.2} 
\end{eqnarray} 
where $\sigma _{ij}=\left( \sigma _{i}+\sigma _{j}\right) /2$, $\widehat{% 
\bbox {\sigma}}$ is a unit vector along their line of centers, $\Theta $ is 
the Heaviside step function, and ${\bf g}_{12}={\bf v}_{1}-{\bf v}_{2}$. The 
primes on the velocities denote the initial values $\{{\bf v}_{1}^{\prime },  
{\bf v}_{2}^{\prime }\}$ that lead to $\{{\bf v}_{1},{\bf v}_{2}\}$ 
following a binary (restituting) collision:  
\begin{equation} 
{\bf v}_{1}^{\prime }={\bf v}_{1}-\mu _{ji}\left( 1+\alpha _{ij}^{-1}\right) 
(\widehat{{\bbox {\sigma }}}\cdot {\bf g}_{12})\widehat{{\bbox {\sigma }}} 
,\quad {\bf v}_{2}^{\prime }={\bf v}_{2}+\mu _{ij}\left( 1+\alpha 
_{ij}^{-1}\right) (\widehat{{\bbox {\sigma }}}\cdot {\bf g}_{12})\widehat{% 
\bbox {\sigma}}  \label{2.3} 
\end{equation} 
where $\mu _{ij}=m_{i}/\left( m_{i}+m_{j}\right) $. The relevant 
hydrodynamic fields are the number densities $n_{i}$, the flow velocity $% 
{\bf u}$, and the temperature $T$. They are defined in terms of moments of 
the distributions $f_{i}$ as  
\begin{equation} 
n_{i}=\int d{\bf v}_{1}f_{i}({\bf v}_{1})\;,\quad \rho {\bf u}=\sum_{i}\int 
d {\bf v}_{1}m_{i}{\bf v}_{1}f_{i}({\bf v}_{1})\;,  \label{2.4} 
\end{equation} 
\begin{equation} 
nT=p=\sum_{i}\int d{\bf v}_{1}\frac{m_{i}}{3}V_{1}^{2}f_{i}({\bf v}_{1})\;, 
\label{2.5} 
\end{equation} 
where ${\bf V}_{1}={\bf v}_{1}-{\bf u}$ is the peculiar velocity, $% 
n=n_{1}+n_{2}$ is the total number density, $\rho =m_{1}n_{1}+m_{2}n_{2}$ is 
the total mass density, and $p$ is the pressure. 
 
The collision operators conserve the particle number of each species and the 
total momentum but the total energy is not conserved:  
\begin{equation} 
\int d{\bf v}_{1}J_{ij}[{\bf v}_{1}|f_{i},f_{j}]=0\;,  \label{2.6} 
\end{equation} 
\begin{equation} 
\sum_{i,j}\int d{\bf v}_{1}m_{i}{\bf v}_{1}J_{ij}[{\bf v}_{1}|f_{i},f_{j}]=0 
\;,  \label{2.7} 
\end{equation} 
\begin{equation} 
\sum_{i,j}\int d{\bf v}_{1}\case{1}{2}m_{i}v_{1}^{2}J_{ij}[{\bf v} 
_{1}|f_{i},f_{j}]=-\case{3}{2}nT\zeta \;,  \label{2.8} 
\end{equation} 
where $\zeta $ is identified as the ``cooling rate'' due to inelastic 
collisions among all species. From Eqs.\ (\ref{2.4})--(\ref{2.8}), the 
macroscopic balance equations for the mixture can be obtained. They are 
given by  
\begin{equation} 
D_{t}n_{i}+n_{i}\nabla \cdot {\bf u}+\frac{\nabla \cdot {\bf j}_{i}}{m_{i}} 
=0\;,  \label{2.9} 
\end{equation} 
\begin{equation} 
D_{t}{\bf u}+\rho ^{-1}\nabla {\sf P}=0\;,  \label{2.10} 
\end{equation} 
\begin{equation} 
D_{t}T-\frac{T}{n}\sum_{i}\frac{\nabla \cdot {\bf j}_{i}}{m_{i}}+\frac{2}{3n} 
\left( \nabla \cdot {\bf q}+{\sf P}:\nabla {\bf u}\right) =-\zeta \,T\;. 
\label{2.11} 
\end{equation} 
In the above equations, $D_{t}=\partial _{t}+{\bf u}\cdot \nabla $ is the 
material derivative,  
\begin{equation} 
{\bf j}_{i}=m_{i}\int d{\bf v}_{1}\,{\bf V}_{1}\,f_{i}({\bf v}_{1})
\label{2.11bb} 
\end{equation} 
is the mass flux for species $i$ relative to the local flow,  
\begin{equation} 
{\sf P}=\sum_{i}\,\int d{\bf v}_{1}\,m_{i}{\bf V}_{1}{\bf V}_{1}\,f_{i}({\bf  
v}_{1})  \label{2.12} 
\end{equation} 
is the total pressure tensor, and  
\begin{equation} 
{\bf q}=\sum_{i}\,\int d{\bf v}_{1}\,\case{1}{2}m_{i}V_{1}^{2}{\bf V} 
_{1}\,f_{i}({\bf v}_{1})  \label{2.13} 
\end{equation} 
is the total heat flux. 
 
The balance equations become a closed set of hydrodynamic equations for the 
fields $n_{i}$, ${\bf u}$ and $T$ once the fluxes (\ref{2.11})--(\ref{2.13}) 
and the cooling rate $\zeta $ are obtained in terms of the hydrodynamic 
fields and their gradients. The resulting equations constitute the 
hydrodynamics for the mixture. Since these fluxes are explicit linear 
functionals of $f_{i}$, a \ representation in terms of the fields results 
when a solution to the Boltzmann equation can be obtained as a function of 
the fields and their gradients. Such a solution is called a ``normal'' 
solution, and a practical means to generate it for small spatial gradients 
is provided by the Chapman-Enskog method. \cite{FK72}

\section{Chapman-Enskog solution of the Boltzmann equations} 
\label{sec3} 
 
The analysis of transport phenomena in fluid mixtures is considerably more 
complicated than in the case of a one component system. Not only is the 
number of transport coefficients larger but these coefficients also are 
functions of more parameters such as the concentrations, mass ratios, size 
ratios, and the three coefficients of restitution. It follows from fluid 
symmetry that the pressure tensor has the same form to first order in the 
gradients as for a one component system. As noted in the Introduction, there 
is more flexibility in the representation of the heat and mass fluxes. Even 
in the case of elastic collisions, several different (but equivalent) 
choices of hdyrodynamic fields are used so some care is required in 
comparing transport coefficients in the different representations. The 
choice here is to use the concentration of species $1$, defined in terms of 
the densities by $x_{1}=n_{1}/\left( n_{1}+n_{2}\right) $, together with the 
pressure, the temperature, and the three components of the local flow 
velocity. \cite{Zubarev} The fluxes then have the forms  
\begin{equation} 
{\bf j}_{1}=-\left( \frac{m_{1}m_{2}n}{\rho }\right) D\nabla x_{1}-\frac{% 
\rho }{p}D_{p}\nabla p-\frac{\rho }{T}D^{\prime }\nabla T,\hspace{0.3in}{\bf % 
j}_{2}=-{\bf j}_{1}  \label{n1} 
\end{equation} 
\begin{equation} 
{\bf q}=-T^{2}D^{\prime \prime }\nabla x_{1}-L\nabla p-\lambda \nabla T, 
\label{n2} 
\end{equation} 
\begin{equation} 
P_{\alpha \beta }=p\delta _{\alpha \beta }-\eta \left( \nabla _{\beta 
}u_{\alpha }+\nabla _{\alpha }u_{\beta }-\frac{2}{3}\delta _{\alpha \beta }% 
{\bf \nabla \cdot u}\right) -\kappa \delta _{\alpha \beta }{\bf \nabla \cdot 
u}.  \label{n3} 
\end{equation} 
The transport coefficients in these equations are  
\begin{equation} 
\left(  
\begin{array}{c} 
D \\  
D^{\prime } \\  
D_{p} \\  
D^{\prime \prime } \\  
\lambda  \\  
L \\  
\eta  \\  
\kappa  
\end{array} 
\right) =\left(  
\begin{array}{c} 
\text{diffusion coefficient} \\  
\text{thermal diffusion coefficient} \\  
\text{pressure diffusion coefficient} \\  
\text{Dufour coefficient} \\  
\text{thermal conductivity} \\  
\text{pressure energy coefficient} \\  
\text{shear viscosity} \\  
\text{bulk viscosity} 
\end{array} 
\right)   \label{n4} 
\end{equation} 
For systems with elastic collisions, the thermal conductivity in a mixture 
is generally measured at ${\bf j}_{1}=$ constant, based on Onsager's 
relations between coefficients in ${\bf j}_{1}$ and ${\bf q}$. This is no 
longer an experimentally useful choice here. If in addition it is required 
that $p=$ constant, then $\nabla x_{1}$ can be eliminated to give ${\bf q}$ 
in terms of ${\bf j}_{1}$ and $\nabla T$. The coefficient of $\nabla T$ in 
this representation is then the thermal conductivity.\cite{GM84} The bulk viscosity 
vanishes at low density, as shown below, just as in the case of elastic 
collisions. The objective here is to apply the Chapman-Enskog method for a 
solution to the Boltzmann equation to first order in the gradients, confirm 
the forms (\ref{n1})--(\ref{n3}), and determine a means to calculate the 
transport coefficients as functions of the parameters of the system. 
 
The Chapman-Enskog method assumes the existence of a ``normal'' solution in 
which all space and time dependence of the distribution function occurs 
through a functional dependence on the hydrodynamic fields  
\begin{equation} 
f_{i}({\bf r},{\bf v}_{1},t)=f_{i}\left[ {\bf v}_{1}|x_{1}
({\bf r}, t), p({\bf r}, t), T({\bf r}, t), {\bf u}({\bf r}, t) \right] \;.  
\label{3.1} 
\end{equation} 
For small spatial variations, this functional dependence can be made local 
in space and time through an expansion in gradients of the fields. To 
generate the expansion $f_{i}$ is written as a series expansion in a formal 
parameter $\epsilon $ measuring the uniformity of the system,  
\begin{equation} 
f_{i}=f_{i}^{(0)}+\epsilon \,f_{i}^{(1)}+\epsilon^2 \,f_{i}^{(2)}+\cdots \;, 
\label{3.2} 
\end{equation} 
where each factor of $\epsilon $ means an implicit gradient of a 
hydrodynamic field. The local reference states $f_{i}^{(0)}$ are chosen such 
that they have the same moments as in Eqs.\ (\ref{2.4}) and (\ref{2.5}), or 
equivalently, the remainder of the expansion must obey the orthogonality 
conditions  
\begin{equation} 
\int d{\bf v}_{1}\left[ f_{i}({\bf v}_{1})-f_{i}^{(0)}({\bf v}_{1})\right] 
=0\;,\quad \sum_{i}\int d{\bf v}_{1}m_{i}{\bf v}_{1}\left[ f_{i}({\bf v}
_{1})-f_{i}^{(0)}({\bf v}_{1})\right] ={\bf 0}\;,  \label{3a} 
\end{equation} 
\begin{equation} 
\sum_{i}\int d{\bf v}_{1}\frac{m_{i}}{2}v_{1}^{2}\left[ f_{i}({\bf v}
_{1})-f_{i}^{(0)}({\bf v}_{1})\right] =0\;,  \label{3b} 
\end{equation} 
The time derivatives of the fields are also expanded as $\partial 
_{t}=\partial _{t}^{(0)}+\epsilon \partial _{t}^{(1)}+\cdots $. The 
coefficients of the time derivative expansion are identified from the 
balance equations (\ref{2.9})--(\ref{2.11}) with a representation of the 
fluxes and the cooling rate in the macroscopic balance equations as a 
similar series through their definitions as functionals of the distributions  
$f_{i}$. This is the usual Chapman-Enskog method for solving kinetic 
equations. \cite{FK72} The main difference in the case of inelastic 
collisions is that the reference state has a time dependence 
associated with the cooling that is not proportional to a spatial gradient. 
Consequently, terms from the time derivative $\partial _{t}^{(0)}$ 
are not zero. 
 
To zeroth order in $\epsilon $, the kinetic equations (\ref{2.1}) become  
\begin{equation} 
\partial_{t}^{(0)}f_{i}^{(0)}=\sum_{j}\,J_{ij}[f_{i}^{(0)},f_{j}^{(0}]\;. 
\label{3.3} 
\end{equation} 
The mass and energy balance equations to this order give  
\begin{equation} 
\partial _{t}^{(0)}x_{i}=0\,\;,\quad T^{-1}\partial 
_{t}^{(0)}T=p^{-1}\partial _{t}^{(0)}p=-\zeta ^{(0)}\;  \label{3.4} 
\end{equation} 
where the cooling rate $\zeta ^{(0)}$ is determined by Eq.\ (\ref{2.8}) to 
zeroth order  
\begin{equation} 
\zeta ^{(0)}=-\frac{2}{3p}\sum_{i,j}\int d{\bf v}_{1}\,\frac{1}{2}
m_{i}\,v_{1}^{2}J_{ij}[{\bf v}_{1}|f_{i}^{(0)},f_{j}^{(0}]\;.  \label{3.5} 
\end{equation} 
This homogeneous state has been studied recently \cite{GD299} and is 
discussed in more detail in the next Section. The time derivative in 
Eq.\ (\ref{3.3}) can be represented more usefully as  
\begin{equation} 
\partial _{t}^{(0)}f_{i}^{(0)}=-\zeta ^{(0)}\left( T\partial _{T}+p\partial 
_{p}\right) f_{i}^{(0)}=\frac{1}{2}\zeta ^{(0)}\nabla _{v_{1}}\cdot \left(  
{\bf V}_{1}f_{i}^{(0)}\right) \;,  \label{3.9} 
\end{equation} 
where $\nabla _{v_{1}}=\partial /\partial {\bf v}_{1}$. The second equality 
follows from dimensional analysis which requires that the temperature 
dependence of $f_{i}^{(0)}$ must be of the form  
\begin{equation} 
f_{i}^{(0)}=x_{i}\frac{p}{T}v_{0}^{-3}\Phi _{i}\left(V_{1}/v_{0}\right) \;, 
\label{3.10} 
\end{equation} 
where $v_{0}^{2}(t)=2T(t)(m_{1}+m_{2})/\left( m_{1}m_{2}\right) $ is a 
thermal velocity defined in terms of the temperature $T(t)$ of the mixture. 
The dependence on the magnitude of ${\bf V}_{1}$ follows from the 
requirement that to zeroth order in gradients $f_{i}$ must be isotropic with 
respect to the peculiar velocity. The Boltzmann equations at this order can be 
written finally as  
\begin{equation} 
\frac{1}{2}\zeta ^{(0)}\nabla _{v_{1}}\cdot \left( {\bf V}
_{1}f_{i}^{(0)}\right) =\sum_{j}\,J_{ij}[f_{i}^{(0)},f_{j}^{(0}]\;. 
\label{3.11} 
\end{equation} 
Since the distribution functions are isotropic, it follows from Eqs.\ (\ref 
{2.11bb}) and (\ref{2.13}), that the zeroth order mass and heat fluxes 
vanish while, for the same reason, the momentum flux is diagonal with a 
coefficient that is just the sum of the partial pressures, i.e.,  
\begin{equation} 
{\bf j}_{i}^{(0)}={\bf 0,}\,\hspace{0.3in}{\bf q}^{(0)}={\bf 0,}\,\hspace{
0.3in}P_{\alpha \beta }^{(0)}=p\delta _{\alpha \beta }.  \label{3.12} 
\end{equation} 
 
To first order in the gradients, the equation for $f_{i}^{(1)}$ is  
\begin{equation} 
\left( \partial _{t}^{(0)}+{\cal L}_{i}\right) f_{i}^{(1)}+{\cal M}
_{i}f_{j}^{(1)}=-\left( \partial _{t}^{(1)}+{\bf v}_{1}\cdot \nabla \right) 
f_{i}^{(0)}\;,  \label{3.13} 
\end{equation} 
where it is understood that $i\neq j$ and the linear operators ${\cal L}_{i}$ 
and ${\cal M}_{i}$ are  
\begin{equation} 
{\cal L}_{i}f_{i}^{(1)}=-\left( 
J_{ii}[f_{i}^{(0)},f_{i}^{(1)}]+J_{ii}[f_{i}^{(1)},f_{i}^{(0)}]+
J_{ij}[f_{i}^{(1)},f_{j}^{(0)}]\right) \;, 
\label{3..14} 
\end{equation} 
\begin{equation} 
{\cal M}_{i}f_{j}^{(1)}=-J_{ij}[f_{i}^{(0)},f_{j}^{(1)}].  \label{3.15} 
\end{equation} 
The action of the time derivatives $\partial _{t}^{(1)}$ on the hydrodynamic 
fields is  
\begin{equation} 
D_{t}^{(1)}x_{1}=0,  \label{3.16bis} 
\end{equation} 
\begin{equation} 
D_{t}^{(1)}p=-\frac{5p}{3}\nabla \cdot {\bf u},  \label{3.16} 
\end{equation} 
\begin{equation} 
D_{t}^{(1)}T=-\frac{2T}{3}\nabla \cdot {\bf u},  \label{3.17} 
\end{equation} 
\begin{equation} 
D_{t}^{(1)}{\bf u}=-\rho ^{-1}{\bf \nabla }p,  \label{3.18} 
\end{equation} 
where $D_{t}^{(1)}=\partial _{t}^{(1)}+{\bf u\cdot \nabla }$ and use has 
been made of the results ${\bf j}_{i}^{(0)}={\bf q}^{(0)}=\zeta ^{(1)}=0$. The 
last equality follows from the fact that the cooling rate is a scalar, and 
corrections to first order in the gradients can arise only from the divergence 
of the vector field. However, as is demonstrated  below, there is no 
contribution to the distribution function proportional to this divergence. 
We note that this is special to the low density Boltzmann equation and 
such terms do occur at higher densities.  
Evaluating the right side of Eq.\ (\ref{3.13}) gives  
\begin{eqnarray} 
-\left( \partial _{t}^{(1)}+{\bf v}_1\cdot \nabla \right) f_{i}^{(0)} 
&=&-\left( \frac{\partial }{\partial x_{1}}f_{i}^{(0)}\right) _{p,T}{\bf V}_1
\cdot \nabla x_{1}-\left[ f_{1}^{(0)}{\bf V}_1+\frac{nT}{\rho }\left( \frac{
\partial }{\partial {\bf V}}_1f_{i}^{(0)}\right) \right] \cdot \nabla \ln p  
\nonumber \\ 
&&+\left[ f_{i}^{(0)}+\frac{1}{2}\frac{\partial}{\partial {\bf V}_1}\cdot 
\left({\bf V}_1f_{i}^{(0)}\right) \right] {\bf V}_1\cdot \nabla \ln T  
\nonumber \\ 
&&+\left( V_{1\alpha}\frac{\partial}{\partial V_{1\beta 
}}f_{i}^{(0)}-\frac{1}{3}\delta _{\alpha \beta }{\bf V}_1\cdot 
\frac{\partial}{\partial {\bf V}_1} f_{i}^{(0)}\right) \nabla _{\alpha 
}u_{\beta }.  \label{3.19} 
\end{eqnarray} 
The equation for $f_{i}^{(1)}$ is now  
\begin{equation} 
\left( \partial _{t}^{(0)}+{\cal L}_{i}\right) f_{i}^{(1)}+{\cal M}
_{i}f_{j}^{(1)}={\bf A}_{i}\cdot \nabla x_{1}+{\bf B}_{i}\cdot \nabla p+{\bf 
C}_{i}\cdot \nabla T+D_{i,\alpha \beta }\nabla _{\alpha }u_{\beta }. 
\label{3.20} 
\end{equation} 
The coefficients of the field gradients on the right side are functions of $
{\bf V}_1$ and the hydrodynamic fields. They are given by  
\begin{equation} 
{\bf A}_{i}({\bf V}_1)=-\left(\frac{\partial}{\partial x_{1}}
f_{i}^{(0)}\right)_{p,T}{\bf V}_1,  \label{3.21} 
\end{equation} 
\begin{equation} 
{\bf B}_{i}({\bf V}_1)=-\frac{1}{p}\left[ f_{i}^{(0)}{\bf 
V}_1+\frac{nT}{\rho }
\left(\frac{\partial}{\partial {\bf V}_1}f_{i}^{(0)}\right) \right] , 
\label{3.22} 
\end{equation} 
\begin{equation} 
{\bf C}_{i}({\bf V}_1)=\frac{1}{T}\left[ 
f_{i}^{(0)}+\frac{1}{2}\frac{\partial }{\partial {\bf V}_1}\cdot \left( {\bf 
V}_1f_{i}^{(0)}\right) \right] {\bf V}_1, \label{3.23} \end{equation} 
\begin{equation} 
D_{i,\alpha \beta }({\bf V}_1)=V_{1\alpha }\frac{\partial }{\partial 
V_{1\beta }}
f_{i}^{(0)}-\frac{1}{3}\delta_{\alpha \beta }{\bf V}_1\cdot \frac{\partial 
}{\partial {\bf V}_1}f_{i}^{(0)}.  \label{3.24} 
\end{equation} 
Note that the trace of $D_{i,\alpha\beta}$ vanishes, confirming that 
the distribution function has not contribution from the divergence of the 
flow field.  The solutions to Eqs.\ (\ref{3.20}) are of the form  
\begin{equation} 
f_{i}^{(1)}={\sf {\cal A}}_{i}\cdot \nabla x_{1}+{\sf {\cal B}}_{i}\cdot 
\nabla p+{\sf {\cal C}}_{i}\cdot \nabla T+{\sf {\cal D}}_{i,\alpha \beta 
}\nabla _{\alpha }u_{\beta }\;.  \label{3.25} 
\end{equation} 
The coefficients ${\sf {\cal A}}_{i},{\sf {\cal B}}_{i},{\sf {\cal C}}_{i},$ 
and ${\sf {\cal D}}_{i,\alpha \beta }$ are functions of the peculiar 
velocity ${\bf V}_1$ and the hydrodynamic fields. The cooling rate 
depends on space through its dependence on $x_{1}$, $p$, and $T$. The time 
derivative $\partial _{t}^{(0)}$ acting on these quantities can be evaluated 
by the replacement $\partial _{t}^{(0)}\rightarrow -\zeta ^{(0)}\left( 
T\partial _{T}+p\partial _{p}\right) $. In addition, there are contributions 
from $\partial _{t}^{(0)}$ acting on the temperature and pressure gradients 
given by  
\begin{eqnarray} 
\partial _{t}^{(0)}\nabla T &=&-\nabla \left( T\zeta ^{(0)}\right) =-\zeta 
^{(0)}\nabla T-T\nabla \zeta ^{(0)}  \nonumber \\ 
&=&-\frac{\zeta ^{(0)}}{2T}\nabla T-T\left[ \left( \frac{\partial \zeta 
^{(0)}}{\partial x_{1}}\right) _{p,T}\nabla x_{1}+\frac{\zeta ^{(0)}}{p}
\nabla p\right] ,  \label{3.26} 
\end{eqnarray} 
\begin{eqnarray} 
\partial _{t}^{(0)}\nabla p &=&-\nabla \left( p\zeta ^{(0)}\right) =-\zeta 
^{(0)}\nabla p-p\nabla \zeta ^{(0)}  \nonumber \\ 
&=&-2\zeta ^{(0)}\nabla p-p\left[ \left( \frac{\partial \zeta ^{(0)}}{
\partial x_{1}}\right) _{p,T}\nabla x_{1}-\frac{\zeta ^{(0)}}{2T}\nabla T
\right] .  \label{3.27} 
\end{eqnarray} 
The integral equations for ${\sf {\cal A}}_{i},{\sf {\cal B}}_{i},{\sf {\cal 
C}}_{i},$ and ${\sf {\cal D}}_{i,\alpha \beta }$ are identified as 
coefficients of the independent gradients in Eq.\ (\ref{3.25}):  
\begin{equation} 
\left[ -\zeta ^{(0)}\left( T\partial _{T}+p\partial _{p}\right) +{\cal L}_{i}
\right] {\sf {\cal A}}_{i}+{\cal M}_{i}{\sf {\cal A}}_{j}={\bf A}_{i}+\left(  
\frac{\partial \zeta ^{(0)}}{\partial x_{1}}\right) _{p,T}\left( p{\sf {\cal 
 B}}_{i}+T{\sf {\cal C}}_{i}\right) ,  \label{3.28} 
\end{equation} 
\begin{equation} 
\left[ -\zeta ^{(0)}\left( T\partial _{T}+p\partial _{p}\right) +{\cal L}
_{i}-2\zeta ^{(0)}\right] {\cal B}_{i}+{\cal M}_{i}{\cal B}_{j}={\bf B}_{i}+ 
\frac{T\zeta ^{(0)}}{p}{\sf {\cal C}}_{i},  \label{3.29} 
\end{equation} 
\begin{equation} 
\left[ -\zeta ^{(0)}\left( T\partial _{T}+p\partial _{p}\right) +{\cal L}
_{i}-\frac{1}{2}\zeta ^{(0)}\right] {\cal C}_{i}+{\cal M}_{i}{\cal C}_{j}= 
{\bf C}_{i}-\frac{p\zeta ^{(0)}}{2T}{\sf {\cal B}}_{i},  \label{3.30} 
\end{equation} 
\begin{equation} 
\left[ -\zeta ^{(0)}\left( T\partial _{T}+p\partial _{p}\right) +{\cal 
L}_{i}\right] {\sf {\cal D}}_{i,\alpha \beta }+{\cal M}_{i}{\sf {\cal D}}
_{j,\alpha \beta }=D_{i,\alpha \beta }.  \label{3.31} 
\end{equation} 
The solutions to these linear integral equations are made unique by the 
orthogonality conditions (\ref{3a}) and (\ref{3b}), i.e.,  
\begin{equation} 
\int d{\bf v}_{1}\left(  
\begin{array}{c} 
{\sf {\cal A}}_{i} \\  
{\sf {\cal B}}_{i} \\  
{\sf {\cal C}}_{i} \\  
{\sf {\cal D}}_{i,\alpha \beta } 
\end{array} 
\right) =0\;,\quad \sum_{i}\int d{\bf v}_{1}m_{i}{\bf V}_{1}\left(  
\begin{array}{c} 
{\sf {\cal A}}_{i} \\  
{\sf {\cal B}}_{i} \\  
{\sf {\cal C}}_{i} \\  
{\sf {\cal D}}_{i,\alpha \beta } 
\end{array} 
\right) ={\bf 0}\;,  \label{3.32} 
\end{equation} 
\begin{equation} 
\sum_{i}\int d{\bf v}_{1}\frac{m_{i}}{2}V_{1}^{2}\left(  
\begin{array}{c} 
{\sf {\cal A}}_{i} \\  
{\sf {\cal B}}_{i} \\  
{\sf {\cal C}}_{i} \\  
{\sf {\cal D}}_{i,\alpha \beta } 
\end{array} 
\right) =0\;.  \label{3.33} 
\end{equation} 
With the functions $\left( {\sf {\cal A}}_{i},{\sf {\cal B}}_{i},{\sf {\cal C }}_{i},{\sf {\cal D}}_{i,\alpha \beta }\right) $ determined in this way, the 
solutions to the Boltzmann equations are determined by (\ref{3.25}) exactly 
to first order in the spatial gradients. 
 
Use of Eq.\ (\ref{3.25}) in the definitions (\ref{2.11bb})--(\ref{2.13}) 
gives the expected forms (\ref{n1})--(\ref{n3}) for the fluxes. The 
transport coefficients associated with ${\bf j}_{1}$ are identified as  
\begin{equation} 
D=-\frac{\rho }{3m_{2}n}\int d{\bf v}_{1}\,{\bf V}_{1}{\bf \cdot \,}{\cal A} 
_{1},  \label{3.34} 
\end{equation} 
\begin{equation} 
D_{p}=-\frac{m_{1}p}{3\rho}\int d{\bf v}_{1}\,{\bf V}_{1}\,\cdot {\sf {\cal B% 
}}_{1} ,  \label{3.35} 
\end{equation} 
\begin{equation} 
D^{\prime }=-\frac{m_{1}T}{3\rho }\int d{\bf v}_{1}\,{\bf V}_{1}\,\cdot {\sf  
{\cal C}}_{1} .  \label{3.36} 
\end{equation} 
The transport coefficients for the heat flux are  
\begin{equation} 
D^{\prime \prime }=-\frac{1}{3T^{2}}\sum_{i}\,\int d{\bf v}_{1}\,\frac{1}{2} 
m_{i}V_{1}^{2}{\bf V}_{1}\,\cdot {\sf {\cal A}}_{i},  \label{3.37} 
\end{equation} 
\begin{equation} 
L=-\frac{1}{3}\sum_{i}\,\int d{\bf v}_{1}\,\frac{1}{2}m_{i}V_{1}^{2}{\bf V} 
_{1}\cdot \,{\sf {\cal B}}_{i},  \label{3.38} 
\end{equation} 
\begin{equation} 
\lambda =-\frac{1}{3}\sum_{i}\,\int d{\bf v}_{1}\,\frac{1}{2}m_{i}V_{1}^{2}  
{\bf V}_{1}\,\cdot {\sf {\cal C}}_{i}.  \label{3.39} 
\end{equation} 
Finally, the shear and bulk viscosities are  
\begin{equation} 
\eta =-\frac{1}{10}\sum_{i}\,\int d{\bf v}_{1}\, m_{i}V_{1\alpha}V_{1\beta}
{\sf {\cal D}}_{i,\alpha\beta},  \label{3.40} 
\end{equation} 
\begin{equation} 
\kappa =-\frac{1}{9}\sum_{i}\,\int d{\bf v}_{1}\, m_{i}V_{1}^{2}
{\sf {\cal D}}_{i,\alpha\alpha}=0.  
\label{3.41} 
\end{equation} 
The bulk viscosity vanishes since the trace of ${\sf {\cal D}}_{i,\alpha 
\beta}$ vanishes, as follows from Eq.\ (\ref{3.31}). 
 
To summarize the results to this point, the solutions to the Boltzmann 
equations to first order in the spatial gradients are  
\begin{equation} 
f_{i}=f_{i}^{(0)}+{\sf {\cal A}}_{i}\cdot \nabla x_{1}+{\sf {\cal B}} 
_{i}\cdot \nabla p+{\sf {\cal C}}_{i}\cdot \nabla T+{\sf {\cal D}}_{i,\alpha 
\beta }\nabla_{\alpha }u_{\beta }\;.  \label{3.42} 
\end{equation} 
The solution to zeroth order, $f_{i}^{(0)}$, is obtained from 
Eq.\ (\ref{3.31}) while the functions ${\sf {\cal A}}_{i},{\sf {\cal 
B}}_{i},{\sf {\cal C}} _{i},$ and ${\sf {\cal D}}_{i,\alpha \beta }$ 
characterizing the solution to first order in the gradients are determined 
from the integral equations (\ref {3.28})--(\ref{3.31}). Calculating the 
mass, heat, and momentum fluxes from this solution one can identify the 
transport coefficients in terms of the integrals (\ref{3.34})--(\ref{3.41}). 
These fluxes, together with the macroscopic balance equations
(\ref{2.9})--(\ref{2.11}), provide the closed set of Navier-Stokes order hydrodynamic 
equations for a granular binary mixture.  All of these results are still 
exact and valid for arbitrary values of the restitution coefficients.

\section{Sonine polynomial approximation} 
\label{sec4} 
 
Accurate approximations to the solutions to the integral equations for $% 
f_{i}^{(0)}$ and $\left( {\sf {\cal A}}_{i},{\sf {\cal B}}_{i},{\sf {\cal C}} 
_{i},{\sf {\cal D}}_{i,\alpha \beta }\right) $ may be obtained using low 
order truncation of expansions in a series of Sonine polynomials. The 
determination of $f_{i}^{(0)}$ to leading order in the Sonine expansion has 
been analyzed elsewhere \cite{GD299} and only the main result is quoted 
here. The polynomials are defined with respect to a Gaussian weight factor. 
The parameters of this Gaussian are chosen such that the leading term in the 
expansion yields the exact moments of the entire distribution with respect 
to $1$, $v_{1}$, and $v_{1}^{2}$. The latter defines the kinetic 
temperatures for each species  
\begin{equation} 
\frac{3}{2}n_{i}T_{i}=\int d{\bf v}_{1}\frac{m_{i}}{2}V_{1}^{2}f_{i}^{(0)}. 
\label{4.0} 
\end{equation} 
For elastic collisions the temperatures $T_{i}$ are the same as the global 
temperature $T$ defined in Eq.\ (\ref{2.5}). The condition that $f_{i}^{(0)}$ 
is normal implies that $T_{i}$ is a function of $n_{i}$ and $T$, or 
equivalently  
\begin{equation} 
\frac{T_{i}}{T}\equiv \gamma _{i}(x_{1}),  \label{4.01} 
\end{equation} 
where $\gamma _{i}$ depends on the hydrodynamic state through the 
concentration. The determination of this functional dependence is worked out 
in Ref.\ \onlinecite{GD299}. The Sonine polynomials used here are defined 
with respect to Maxwellians characterized by the temperatures $T_{i}$ rather 
than those characterized by $T$ \ for elastic collisions. With this choice 
the leading deviation from the Maxwellians is a polynomial of degree $4$  
\begin{equation} 
f_{i}^{(0)}=n_{i}v_{0}^{-3}\Phi _{i}\left( V_{1}^{\ast}\right) \;, 
\label{4.02} 
\end{equation} 
\begin{equation} 
\Phi _{i}(V_{1}^{\ast })\rightarrow \left( \frac{\theta_{i}}{\pi }\right) 
^{3/2}e^{-\theta _{i}V_{1}^{\ast 2}}\left[ 1+\frac{c_{i}}{4}\left( \theta 
_{i}^{2}V_{1}^{\ast 4}-5\theta _{i}V_{1}^{\ast 2}+\frac{15}{4}\right) 
\right] \;,  \label{4.1} 
\end{equation} 
where $V_{1}^{\ast }=V_{1}/v_{0}$, and $\theta _{i}=(\mu _{ji}\gamma 
_{i})^{-1}$. If polynomials defined in terms of $T$ are used the leading 
correction is a polynomial of degree $2$ in $v_{1}^{\ast }$, proportional to  
$T-T_{i}$. The choice of polynomials defined in terms of $T_{i}$ effectively 
resums an infinite set of terms in this second type of expansion. Since $
\gamma _{i}$ is a function of the concentration, a significant new 
contribution to the parameters of the integral equations for the transport 
coefficients occurs through the additional concentration dependence 
associated with temperature difference of the two species. The coefficients 
$c_{i}$ in Eq.\ (\ref{4.1}) are determined by substitution of Eq.\ 
(\ref{4.02}) into Eq.\ (\ref{3.9}) and retaining all terms linear in $c_{i}$ 
for the leading Sonine polynomial approximation. The reader is referred to 
Ref.\ \onlinecite{GD299} for further details. 
 
In the remainder of this section, we will compute the mass flux, the pressure tensor, and the heat flux in the leading Sonine polynomial approximations. Let us consider each one separately.

\subsection{Mass flux}
\label{sub4.1}

Here, the lowest order Sonine polynomial 
approximations for ${\sf {\cal A}}_{i},{\sf {\cal B}}_{i},$ and ${\sf {\cal 
C}}_{i}$ are obtained and applied to the calculation of the transport 
coefficients in the mass fluxes, $D$, $D_{p}$, and $D^{\prime }$. The 
leading Sonine approximations (lowest degree polynomial) of the quantities $
{\sf {\cal A}}_{i}$, ${\sf {\cal B}}_{i}$, and ${\sf {\cal C}}_{i}$ are  
\begin{equation} 
\left(  
\begin{array}{c} 
{\sf {\cal A}}_{i} \\  
{\sf {\cal B}}_{i} \\  
{\sf {\cal C}}_{i} 
\end{array} 
\right) \rightarrow f_{i,M}{\bf V}_1\left(  
\begin{array}{c} 
a_{i,1} \\  
b_{i,1} \\  
c_{i,1} 
\end{array} 
\right) .  \label{4.2} 
\end{equation} 
The coefficients $\{a_{i,1},b_{i,1},c_{i,1}\}$ are related to the transport 
coefficients in this approximation through (\ref{3.34})--(\ref{3.36})  
\begin{equation} 
a_{1,1}=-\frac{1}{\delta \gamma }a_{2,1}=-\frac{m_{1}m_{2}n}{\rho n_{1}T_{1}}D, 
\label{4.3} 
\end{equation} 
\begin{equation} 
b_{1,1}=-\frac{1}{\delta \gamma }b_{2,1}=-\frac{\rho }{pn_{1}T_{1}}D_{p}, 
\label{4.4} 
\end{equation} 
\begin{equation} 
c_{1,1}=-\frac{1}{\delta \gamma }c_{2,1}=-\frac{\rho }{Tn_{1}T_{1}}D^{\prime }. 
\label{4.5} 
\end{equation} 
Here, $\gamma =T_{1}/T_{2}$ and $\delta =n_{1}/n_{2}$. The coefficients $
\{a_{1,1};b_{1,1};c_{1,1}\}$ are determined by substitution of Eq.\ (\ref{4.2}) 
into the integral equations (\ref{3.28})--(\ref{3.30}). The details are 
carried out in Appendix \ref{appA}  with the results  
\begin{equation} 
a_{1,1}=-\left(\nu -\frac{1}{2}\zeta ^{(0)}\right) ^{-1}\left[ \left(\frac{ 
\partial }{\partial x_{1}}\ln n_{1}T_{1}\right) _{p,T}-\left( \frac{\partial 
\zeta ^{(0)}}{\partial x_{1}}\right) _{p,T}\left( pb_{1,1}+Tc_{1,1}\right)  
\right] ,  \label{4.6} 
\end{equation} 
\begin{equation} 
b_{1,1}=-\frac{1}{p}\left( 1-\frac{m_{1}nT}{\rho T_{1}}\right) \left( \nu -
\frac{3}{2}\zeta ^{(0)}+\frac{\zeta ^{(0)2}}{2\nu }\right)^{-1}, 
\label{4.7} 
\end{equation} 
\begin{equation} 
c_{1,1}=-\frac{p\zeta ^{(0)}}{2T\nu }b_{1,1},  \label{4.8} 
\end{equation} 
where the collision frequency $\nu$ is given by  
\begin{equation} 
\nu =-\frac{m_{1}}{3n_{1}T_{1}}\int d{\bf v}_{1}{\bf V}_{1}\cdot \left( 
J_{12}[{\bf v}_{1}|f_{1,M}{\bf V}_{1},f_{2}^{(0)}]-\delta \gamma J_{12}[{\bf 
v}_{1}|f_{1}^{(0)},f_{2,M}{\bf V}_{2}]\right) .  \label{4.9} 
\end{equation} 
This integral is evaluated in Appendix \ref{appB} with the result  
\begin{eqnarray} 
\frac{\nu}{\nu_{e}} &=&\frac{\left(1+\alpha_{12}\right)}{2}\sqrt{\mu 
_{21}\gamma _{1}+\mu _{12}\gamma _{2}}  \nonumber \\ 
&&\times \left[ 1-\frac{1}{16\rho }\frac{\rho _{1}\left( m_{2}\gamma 
_{1}\right) ^{2}c_{1}+\rho _{2}\left( m_{1}\gamma _{2}\right) ^{2}c_{2}}{
(m_{2}\gamma _{1}+m_{1}\gamma _{2})^{2}}\right] ,  \label{4.10} 
\end{eqnarray} 
where $\nu _{e}$ is the corresponding result for the elastic case  
\begin{equation} 
\nu _{e}=\frac{8}{3}\sqrt{2\pi }\sigma _{12}^{2}\sqrt{\frac{T}{m_{1}m_{2}}}
\frac{\rho }{\sqrt{m_{1}+m_{2}}},  \label{4.11} 
\end{equation} 
and $\rho _{i}=m_{i}n_{i}$ is the mass density of species $i$. These 
results, together with $a_{2,1}=-\delta \gamma a_{1,1},$ $b_{2,1}=-\delta \gamma b_{1,1},$ and $c_{2,1}=-\delta \gamma c_{1,1}$ completely determine the distribution functions to first order in the Sonine polynomial expansion. 
 
The transport coefficients are identified from (\ref{4.6})--(\ref{4.8}) as  
\begin{equation} 
D=\frac{\rho }{m_{1}m_{2}n}\left( \nu -\frac{1}{2}\zeta ^{(0)}\right) ^{-1}
\left[ \left( \frac{\partial }{\partial x_{1}}n_{1}T_{1}\right) _{p,T}+\rho 
\left( \frac{\partial \zeta ^{(0)}}{\partial x_{1}}\right) _{p,T}\left( 
D_{p}+D^{\prime }\right) \right] ,  \label{4.12} 
\end{equation} 
\begin{equation} 
D_{p}=\frac{n_{1}T_{1}}{\rho }\left( 1-\frac{m_{1}nT}{\rho T_{1}}\right) 
\left( \nu -\frac{3}{2}\zeta ^{(0)}+\frac{\zeta ^{(0)2}}{2\nu }\right) ^{-1}, 
\label{4.13} 
\end{equation} 
\begin{equation} 
D^{\prime }=-\frac{\zeta ^{(0)}}{2\nu }D_{p}.  \label{4.14} 
\end{equation} 
Since ${\bf j}_{1}=-{\bf j}_{2}$ and $\nabla x_{1}=-\nabla x_{2}$, it is 
expected that  $D$ should be symmetric with respect to interchange of 
particles $1$ and $2$ while $D_{p}$ and $D^{\prime}$ should be 
antisymmetric. This can be verified by noting that 
$n_{1}T_{1}+n_{2}T_{2}=nT$. The expression for $\nu$ agrees with the known 
result for elastic collisions. \cite{FK72} For the case of mechanically 
equivalent particles ($ 
m_{1}=m_{2},\alpha _{11}=\alpha _{22}=\alpha _{12}\equiv \alpha ,\sigma 
_{11}=\sigma _{22}$) in the dilute concentration limit ($\rho 
_{2}\rightarrow 0$), the expression for the diffusion coefficient $D$ 
coincides with the one recently derived in the self-diffusion 
problem. \cite{BRCG99} 
 
\subsection{Pressure tensor}
\label{sub4.2}

The leading Sonine approximation for the function ${\cal D}_{i,\alpha \beta }$ is 
\begin{equation}
\label{4.2:1}
{\cal D}_{i,\alpha \beta}\to f_{1,M} d_{i,1} R_{i,\alpha \beta}({\bf V}_1),
\end{equation}
where 
\begin{equation}
\label{4.2:2}
R_{i,\alpha \beta}({\bf V}_1)=m_i\left( V_{1\alpha}V_{1\beta}-
\frac{1}{3}V_1^2\delta_{\alpha\beta}\right),
\end{equation}
and 
\begin{equation}
\label{4.2:3}
d_{i,1}=\frac{1}{10}\frac{1}{n_iT_i^2}\int d{\bf v}_1 R_{i,\alpha \beta} {\cal 
D}_{i,\alpha \beta}.
\end{equation}
The coefficients $d_{i,1}$ are related to the shear viscosity $\eta$ in this 
approximation through Eq.\ (\ref{3.40}) as  
\begin{equation}
\label{4.2:4}
\eta=-T^2\left(n_1\gamma_1^2 d_{1,1}+n_2\gamma_2^2 d_{2,1}\right).
\end{equation}

The integral equations for the coefficients $d_{i,1}$ are decoupled from the 
remaining transport coefficients.  The two coefficients 
$d_{i,1}$ are obtained by multiplying Eqs.\  (\ref{3.31}) with $R_{i,\alpha \beta}$ 
and integrating over the velocity to get the coupled set of 
equations
\begin{equation}
\label{4.2:5}
\left(
\begin{array}{cc}
\tau_{11}-\frac{1}{2}\zeta^{(0)}& \tau_{12}\\
\tau_{21}&\tau_{22}-\frac{1}{2}\zeta^{(0)}
\end{array}
\right)
\left(
\begin{array}{c}
d_{1,1}\\
d_{2,1}
\end{array}
\right)
=-\left(
\begin{array}{c}
T_1^{-1}\\
T_2^{-1}
\end{array}
\right).
\end{equation}
The frequencies $\tau_{ij}$ are given in terms of the collision operator by 
\begin{equation}
\label{4.2:6}
\tau_{ii}=\frac{1}{10}\frac{1}{n_iT_i^2}\int d{\bf v}_1 R_{i,\alpha \beta}
{\cal L}_i\left(f_{i,M}R_{i,\alpha \beta}\right), 
\end{equation}
\begin{equation}
\label{4.2:7}
\tau_{ij}=\frac{1}{10}\frac{1}{n_iT_i^2}\int d{\bf v}_1 R_{i,\alpha \beta}
{\cal M}_i\left(f_{j,M}R_{j,\alpha \beta}\right)
, \quad i\neq j.
\end{equation}
The evaluation of these collision integrals is given in Appendix \ref{appC}. The solution with the 
matrix elements known is elementary so that $\eta$ can be calculated directly from 
Eq.\ (\ref{4.2:4}).

\subsection{Heat flux}
\label{sub4.3}

The computation of the heat flux requires going to the second Sonine 
approximation. In this case, the quantities ${\sf {\cal A}}_{i}$, ${\sf {\cal 
B}}_{i}$ and ${\sf {\cal C}}_{i}$ are taken to be
\begin{equation}
\label{4.3:1}
{\sf {\cal A}}_{1}({\bf V}_1)\to f_{1,M}\left[-\frac{m_1m_2n}{\rho 
n_1T_1}D{\bf V}_1+a_{1,2}{\bf S}_1({\bf V}_1)
\right] ,\quad
{\sf {\cal A}}_{2}({\bf V}_1)\to f_{2,M}\left[\frac{m_1m_2n}{\rho 
n_2T_2}D{\bf V}_1+a_{2,2}{\bf S}_2({\bf V}_1)\right] 
\end{equation}
\begin{equation}
\label{4.3:2}
{\sf {\cal B}}_{1}({\bf V}_1)\to f_{1,M}\left[-\frac{\rho}{p 
n_1T_1}D_p{\bf V}_1+b_{1,2}{\bf S}_1({\bf V}_1)
\right] ,\quad
{\sf {\cal B}}_{2}({\bf V}_1)\to f_{2,M}\left[\frac{\rho}{p 
n_2T_2}D_p{\bf V}_1+b_{2,2}{\bf S}_2({\bf V}_1)\right] 
\end{equation}
\begin{equation}
\label{4.3:4}
{\sf {\cal C}}_{1}({\bf V}_1)\to f_{1,M}\left[-\frac{\rho}{T 
n_1T_1}D'{\bf V}_1+c_{1,2}{\bf S}_1({\bf V}_1)
\right] ,\quad
{\sf {\cal C}}_{2}({\bf V}_1)\to f_{2,M}\left[\frac{\rho}{T 
n_2T_2}D'{\bf V}_1+c_{2,2}{\bf S}_2({\bf V}_1)\right],  
\end{equation}
where 
\begin{equation}
\label{4.3:5}
{\bf S}_i({\bf V}_1)=\left(\frac{1}{2}m_iV_1^2-\frac{5}{2}T_i\right){\bf V}_1.
\end{equation} 
In these equations, it is understood that $D$, $D_p$ and $D'$ 
are given by Eqs.\ (\ref{4.12})--(\ref{4.14}), respectively. 
The coefficients $a_{i,2}$, $b_{i,2}$ and $c_{i,2}$ are defined as
\begin{equation}
\left(
\begin{array}{c} 
a_{i,2} \\  
b_{i,2} \\  
c_{i,2}
\end{array}
\right) =\frac{2}{15}\frac{m_i}{n_iT_i^3}
\int d{\bf v}_{1}{\bf S}_i({\bf V}_1)\cdot\left(  
\begin{array}{c} 
{\sf {\cal A}}_{i} \\  
{\sf {\cal B}}_{i} \\  
{\sf {\cal C}}_{i} 
\end{array} 
\right)
\label{4.3:7} 
\end{equation} 
Consequently, the transport coefficients appearing in the heat flux, $D''$, $L$ and 
$\lambda$ are given by 
\begin{equation}
\label{4.3:8}
D''=-\frac{5}{2}T\left(\frac{n_1\gamma_1^3}{m_1}a_{1,2}+
\frac{n_2\gamma_2^3}{m_2}a_{2,2}\right)+\frac{5}{2}\frac{nm_1m_2}{\rho T}
\left(\frac{\gamma_1}{m_1}-\frac{\gamma_2}{m_2}\right)D,
\end{equation}
\begin{equation}
\label{4.3:9}
L=-\frac{5}{2}T^3\left(\frac{n_1\gamma_1^3}{m_1}b_{1,2}+
\frac{n_2\gamma_2^3}{m_2}b_{2,2}\right)+\frac{5}{2}\frac{\rho}{n}
\left(\frac{\gamma_1}{m_1}-\frac{\gamma_2}{m_2}\right)D_p,
\end{equation}
\begin{equation}
\label{4.3:10}
\lambda=-\frac{5}{2}T^3\left(\frac{n_1\gamma_1^3}{m_1}c_{1,2}+
\frac{n_2\gamma_2^3}{m_2}c_{2,2}\right)+\frac{5}{2}\rho
\left(\frac{\gamma_1}{m_1}-\frac{\gamma_2}{m_2}\right)D'.
\end{equation}

The computation of the coefficients $D''$, $L$, and $\lambda$ is also carried out in Appendix \ref{appA}.  By using matrix notation, the coupled set of six equations for the unknowns
\begin{equation}
\label{4.3:11}
\{a_{1,2}; a_{2,2}; b_{1,2}; b_{2,2}; c_{1,2}; c_{2,2}\}
\end{equation}
can be written as 
\begin{equation}
\label{4.3:12}
\Lambda_{\sigma \sigma'}X_{\sigma'}=Y_{\sigma}.
\end{equation}
Here, $X_{\sigma'}$ is the column matrix defined by the set 
(\ref{4.3:11})  and $\Lambda_{\sigma \sigma'}$ is the matrix
\begin{equation}
\label{4.3:13}
\Lambda=\left(
\begin{array} {cccccc}
\nu_{11}-\frac{3}{2}\zeta^{(0)}& \nu_{12}& 
-p\left(\frac{\partial \zeta ^{(0)}}{\partial x_{1}}\right)_{p,T}&0& 
-T\left( \frac{\partial \zeta ^{(0)}}{\partial x_{1}}\right)_{p,T}&0 \\
\nu_{21}&\nu_{22}-\frac{3}{2}\zeta^{(0)}&0&  
p\left( \frac{\partial \zeta ^{(0)}}{\partial x_{1}}\right)_{p,T}&0& 
T\left( \frac{\partial \zeta ^{(0)}}{\partial x_{1}}\right)_{p,T}\\
0& 0& \nu_{11}-\frac{5}{2}\zeta^{(0)}& \nu_{12}& 
-T\zeta^{(0)}/p&0\\ 
0& 0 & \nu_{21} & \nu_{22}-\frac{5}{2}\zeta^{(0)} & 0 &  
-T\zeta^{(0)}/p\\
0& 0& p\zeta^{(0)}/2T&0&\nu_{11}-\zeta^{(0)}& \nu_{12}\\
0& 0& 0&p\zeta^{(0)}/2T&\nu_{21}&\nu_{22}-\zeta^{(0)}
\end{array}
\right).
\end{equation}
Here, we have introduced the collision frequencies
\begin{equation}
\label{4.3:13a}
\nu_{ii}=\frac{2}{15}\frac{m_i}{n_iT_i^3}\int d{\bf v}_1 {\bf S}_i
\cdot {\cal L}_i\left(f_{i,M}{\bf S}_i\right), 
\end{equation}
\begin{equation}
\label{4.3:13b}
\nu_{ij}=\frac{2}{15}\frac{m_i}{n_iT_i^3}\int d{\bf v}_1 {\bf S}_i
\cdot {\cal M}_i\left(f_{j,M}{\bf S}_j\right), \quad i\neq j.
\end{equation}
The column matrix ${\bf Y}$ is
\begin{equation}
\label{4.3:14}
{\bf Y}=\left(
\begin{array}{c}
Y_1\\
Y_2\\
Y_3\\
Y_4\\
Y_5\\
Y_6
\end{array}
\right),
\end{equation}
where
\begin{eqnarray}
\label{4.3:15}
Y_1&=& -\frac{\zeta^{(0)}m_1m_2nD}{\rho 
n_1T_1^2}-\frac{1}{2}\frac{T^2}{n_1T_1^3}\frac{\partial}{\partial x_1}\left(
n_1\gamma_1^2c_1\right)+\frac{2}{15}\frac{m_1^2m_2nD}{\rho n_1^2T_1^4}
\left[\int d{\bf v}_1 {\bf S}_1\cdot {\cal L}_1(f_{1,M}{\bf V}_1) \right.
\nonumber\\
& & \left. -\delta \gamma 
\int d{\bf v}_1 {\bf S}_1\cdot {\cal M}_1(f_{2,M}{\bf V}_2)\right],
\end{eqnarray}
\begin{eqnarray}
\label{4.3:16}
Y_3&=&-\frac{\zeta^{(0)}\rho D_p}{p 
n_1T_1^2}-\frac{1}{2}\frac{c_1}{pT_1}
+\frac{2}{15}\frac{m_1\rho D_p}{p n_1^2T_1^4}
\left[\int d{\bf v}_1 {\bf S}_1\cdot {\cal L}_1(f_{1,M}{\bf V}_1) \right.
\nonumber\\
& & \left. -\delta \gamma 
\int d{\bf v}_1 {\bf S}_1\cdot {\cal M}_1(f_{2,M}{\bf V}_2)\right],
\end{eqnarray}
\begin{eqnarray}
\label{4.3:17}
Y_5&=& -\frac{\zeta^{(0)}\rho D'}{T 
n_1T_1^2}-\frac{2+c_1}{2TT_1}
+\frac{2}{15}\frac{m_1\rho D'}{T n_1^2T_1^4}
\left[\int d{\bf v}_1 {\bf S}_1\cdot {\cal L}_1(f_{1,M}{\bf V}_1) \right.
\nonumber\\
& & \left. -\delta \gamma 
\int d{\bf v}_1 {\bf S}_1\cdot {\cal M}_1(f_{2,M}{\bf V}_2)\right],
\end{eqnarray}
where 
\begin{equation}
\label{4.3:18}
c_i=\frac{8}{15}\left[\frac{m_i^2}{4n_iT_i^2}\int d{\bf v}_1 V_1^4 
f_i^{(0)}-\frac{15}{4}\right].
\end{equation}
The corresponding expressions of the elements $Y_2$, $Y_4$ and $Y_6$ can be 
deduced from Eqs.\ (\ref{4.3:15}), (\ref{4.3:16}) and (\ref{4.3:17}), respectively, by 
interchanging $1\leftrightarrow2$ and setting $D\to D$, $D_p\to -D_p$ and $D'\to -D'$. 
The evaluation of the collision integrals (\ref{4.3:13a}) and (\ref{4.3:13b}) is given in 
Appendix \ref{appD}. 

The solution to Eq.\ (\ref{4.3:12}) is 
\begin{equation}
\label{4.3:19}
X_{\sigma}=\left(\Lambda^{-1}\right)_{\sigma \sigma'}Y_{\sigma'}.
\end{equation}
This relation provides an explicit expression for the coefficients 
$a_{i,2}$, $b_{i,2}$ and $c_{i,2}$ in terms of the restitution coefficients 
and the parameters of the mixture. From these expressions one can easily get 
the transport coefficients $D'$, $L$ and $\lambda$ from 
Eqs.\  (\ref{4.3:8})--(\ref{4.3:10}), respectively.

\section{Discussion} 
 \label{sec5} 
 
The primary objective of this work has been to obtain the hydrodynamic 
description of a binary mixture of granular gases from an underlying kinetic 
theory. The derivation of the hydrodynamic equations consists of two steps. 
First, the macroscopic balance equations (\ref{2.9})--(\ref{2.11}) for 
number densities, total momentum and energy are obtained from the 
corresponding kinetic equation. Next, the fluxes and cooling rate in these 
equations are determined from a solution of the coupled Boltzmann equations 
given in terms of the hydrodynamic fields and their spatial gradients. The 
corresponding constitutive equations for the mass, heat and momentum fluxes 
are given to Navier-Stokes order by Eqs.\ (\ref{n1})--(\ref{n3}), and the 
associated transport coefficients are given by Eqs.\ (\ref{3.34})--(\ref 
{3.36}), (\ref{3.37})--(\ref{3.39}), and (\ref{3.40})--(\ref{3.41}), 
respectively. These results are exact within the context of the Boltzmann 
kinetic equation. A practical evaluation of these coefficients is possible by means of a Sonine polynomial approximation and  
the derivation and approximate results are {\em not} limited to weak inelasticity. An 
exploration of the full parameter space (mass ratio, diameters, 
concentrations, inelasticity parameters) is straightforward but beyond the scope of this presentation.  
We intend to create a library of programs for 
calculation of distribution functions and transport coefficients in one and two component granular systems, available to the public on our web site. 
 
The Chapman-Enskog method provides an expansion of the distribution function for
weak spatial inhomogeneity. This means that the relative spatial variation
of all hydrodynamic fields must be small over distances of order of the mean
free path. This encompasses a wide range of interesting phenomena, but
excludes strongly driven systems such as those under steady shear. For such
states it is not profitable to go beyond the order considered here in the
Chapman-Enskog expansion, but rather to use other methods not based on
small gradients. Previous applications of the Chapman-Enskog method have
typically introduced additional assumptions for convenience that are not
internally consistent with constructing a solution to the Boltzmann equation.
In most of these cases the reference state was chosen to be a Maxwellian,
presumed to give accurate results at weak dissipation. This is not the case,
as has been demonstrated in the one component case.\cite{BDKS98} In addition,
the effects of cooling in the lowest order time derivative of the temperature
and pressure were neglected, again under the assumption of weak dissipation.
Here, the reference state has been taken to be an exact solution to the
uniform Boltzmann equation and consequently, there is no limitation
on the parameters of the system
(mass ratio, size ratio, concentrations, restitution coefficients).
An interesting and important result of this exact analysis is that the
temperatures of each species must be different. Furthermore, the time derivatives
are calculated at each order in the gradient expansion without restriction on
the restitution coefficients. This assures that a consistent solution to the Boltzmann
equation is constructed at each order in the gradients, without any restrictions
on the system parameters. This consistency is reflected in the verification
of solubility conditions for the integral equations determining the transport
coefficients. These are the primary new features of this work.

The evaluation of the transport coefficients for practical results introduces
a new approximation, truncation of an expansion for the solutions to the
integral equations in polynomials. The leading order truncation is known to
be accurate to approximately five percent in the case of elastic collisions.
Exceptions are extreme mass ratios (e.g., electron-proton systems). Its
validity for inelastic systems has been recently checked by Monte Carlo
simulation for shear viscosity. The accuracy is found to be
similar to that for elastic collisions.

Recently, a seemingly similar analysis for granular binary mixtures has been
given in Ref.\ \onlinecite{H00}, also based on a two temperature description.
However, this work is phenomenological with no attempt to solve the kinetic
equation. Instead, it is assumed that the distribution function is a local
Maxwellian. This is reasonably for estimating the dense gas collisional transfer
contributions to the fluxes, but it predicts that all transport coefficients
calculated here at low density should vanish. Clearly, the phenomenology is
flawed.

In a next paper, we will evaluate the expressions for the different transport coefficients for a variety of mass and diameter ratios. Here, as an  
illlustration and to give some insight into the influence of dissipation on transport, 
we consider the pressure diffusion coefficient $D_p$ and the 
diffusion coefficient $D$. For the sake of concreteness, consider the case 
$\alpha _{11}=\alpha _{22}=\alpha _{12}\equiv \alpha $. Figure \ref{fig1} 
shows the reduced pressure diffusion coefficient $D_{p}(\alpha )/D(1)$ as a 
function of the restitution coefficient for $\sigma _{11}=\sigma _{22}$, $
n_{1}/n_{2}=0.25$, and for two values of the mass ratio 
($m_{1}/m_{2}=0.5 \ \text{and}\ 4$). Here, $D_{p}(1)=(n_{1}T/\nu _{e}\rho 
)(1-m_{1}n/\rho )$ is the pressure diffusion coefficient for elastic 
collisions. We see that the deviation from the functional form for elastic 
collisions is quite important even for moderate dissipation (say $\alpha 
\simeq 0.9$). This tendency becomes more significant as the mass of the 
defect particles is larger than that of the excess particles. Also shown for 
comparison in this Fig.\ \ref {fig1} is the result for $m_{1}/m_{2}=4$ with 
$T_{1}=T_{2}=1$, which would be obtained if the differences in the partial 
temperatures were neglected. Clearly, inclusion of this effect makes a 
significant difference over the whole range of dissipation shown (the actual 
value is $\gamma =T_{1}/T_{2}\simeq 1.36$ for $\alpha =0.8$). At the level 
of the mass flux, the main transport coefficient is the diffusion 
coefficient $D$. In Fig.\  \ref{fig2} we plot the dependence of the ratio 
$D(\alpha )/D(1)$ on $\alpha $ for $\sigma _{11}=\sigma _{22}$, 
$n_{1}/n_{2}=0.25$, and for several values of the mass ratio 
($m_{1}/m_{2}=0.5,1,\text{and }4$). As before, $D(1)=(\rho T/m_{1}m_{2}\nu 
_{e})$ is the diffusion coefficient for elastic collisions. The shape of 
these curves is very similar to those presented for the coefficient $D_{p}$, 
although the influence of dissipation on $D$ is a bit stronger than the one 
observed in the case of $D_{p}$. According to the behavior of $D$ and 
$D_{p}$ one can conclude that the main effect of inelasticity in collisions 
is to enhance the mass transport with respect to the elastic collisions 
case.

The application of the Chapman-Enskog procedure here follows closely recent 
derivations of hydrodynamics and transport coefficients for one component 
granular gases. \cite{BDKS98,GD199} An important difference in the mixture 
is the need for two different temperatures in the reference local 
homogeneous cooling states, leading to qualitative differences in the 
concentration dependence of the transport coefficients relative to the 
elastic limit. This reference state is discussed in more detail in Ref.\ 
\onlinecite{GD299}. Effectively, the absence of energy conservation for 
granular gases leads to a failure of detailed balance between the velocity 
distributions in the homogeneous state. One consequence is non Maxwellian 
distributions observed already for one component systems. For multicomponent 
systems a second consequence is different covariances of the distributions 
for different species, although the cooling rates are all equal. The latter 
property implies that the different partial temperatures can be expressed in 
terms of the global temperature (as is required for a hydrodynamic 
description) although the functional relationships defining such partial 
temperatures introduce a new dependence on the concentrations. This leads to 
additional spatial gradients at first order in the Chapman-Enskog expansion 
and consequently additional contributions to the transport coefficients. The 
effect of these terms can be significant, as illustrated in Fig. \ref{fig1}. 
Previous work on granular mixtures \cite{JM89,Z95,AW98,WA99} is restricted 
by weak dissipation approximations. In addition, these studies do not 
include contributions from the partial temperature differences which are significant even for weak dissipation. 
 
One interesting question is whether the mixture hydrodynamics is more or less stable than 
that of the one component case, and if such phenomena as phase separation or 
segregation can occur. Another direction of study is the extension of the 
present simple hydrodynamic state to higher densities based on the revised 
Enskog kinetic equation, following recent results for the one component 
fluid. \cite{GD199} In this case the new complexity is due to the 
dependence of the collision operator on concentration through the pair 
correlation functions for the different species. Finally, we hope that the 
present results stimulate the performance of computer simulations (molecular 
dynamics and/or by using the Direct Simulation Monte Carlo method 
\cite{Bird}) to study hydrodynamics in granular binary mixtures.

\acknowledgements 
We are grateful to Hector S\'anchez-Pajares for helping us in the computation of 
some of the collision integrals. This research was supported by National Science Foundation Grant PHY 
9722133. V. G. also acknowledges partial support from the Ministerio de 
Ciencia y Tecnolog\'{\i}a (Spain)  through Grant No. BFM2001-0718.

\appendix 
\section{Leading Sonine Approximations} 
\label{appA} 
 
In this Appendix the coefficients  $\{a_{i,j};b_{i,j};c_{i,j}\}$ in the leading  Sonine approximations are evaluated.  First, let us consider the 
coefficients $\{a_{1,1};b_{1,1};c_{1,1}\}$ determining the mass flux.  
Substitution of Eq.\ (\ref{4.2}) 
into the integral equations (\ref{3.28})--(\ref{3.30}) gives  
\begin{eqnarray} 
-\zeta ^{(0)}\left( T\partial _{T}+p\partial _{p}\right) a_{1,1}f_{1,M}{\bf V}
_{1} &+&a_{1}\left[ {\cal L}_{1}f_{1,M}{\bf V}_{1}-\delta \gamma {\cal M} 
_{1}f_{2,M}{\bf V}_{1}\right]   \nonumber \\ 
&=&{\bf A}_{1}+\left( \frac{\partial \zeta ^{(0)}}{\partial x_{1}}\right) 
_{p,T}f_{1,M}\left( pb_{1,1}+Tc_{1,1}\right) {\bf V}_{1},  \label{b.1} 
\end{eqnarray} 
\begin{eqnarray} 
\left[-\zeta ^{(0)}\left( T\partial _{T}+p\partial _{p}\right) -2\zeta 
^{(0)}\right] b_{1,1}f_{1,M}{\bf V}_{1} &+&b_{1,1}\left[ {\cal L}_{1}f_{1,M}{\bf 
V}_{1}-\delta \gamma {\cal M}_{1}f_{2,M}{\bf V}_{1}\right]   \nonumber \\ 
&=&{\bf B}_{1}+\frac{T\zeta ^{(0)}}{p}f_{1,M}c_{1,1}{\bf V}_{1},  \label{b.2} 
\end{eqnarray} 
\begin{eqnarray} 
\left[ -\zeta ^{(0)}\left( T\partial _{T}+p\partial _{p}\right) -\frac{1}{2}
\zeta ^{(0)}\right] c_{1,1}f_{1,M}{\bf V}_{1} &+&c_{1,1}\left[ {\cal L}
_{1}f_{1,M}{\bf V}_{1}-\delta \gamma {\cal M}_{1}f_{2,M}{\bf V}_{1}\right]   
\nonumber \\ 
&=&{\bf C}_{1}-\frac{p\zeta ^{(0)}}{2T}f_{1,M}b_{1,1}{\bf V}_{1}.  \label{b.3} 
\end{eqnarray} 
Next, multiplication by $m_{1}{\bf V}_{1}$ and integrating over the velocity leads to  
\begin{equation} 
\left[ -\zeta ^{(0)}\left( T\partial _{T}+p\partial _{p}\right) +\nu \right]
n_{1}T_{1}a_{1,1}=\frac{1}{3}\int d{\bf V}_{1}m_{1}{\bf V}_{1}\cdot {\bf A}
_{1}+\left( \frac{\partial \zeta ^{(0)}}{\partial x_{1}}\right) 
_{p,T}n_{1}T_{1}\left( pb_{1,1}+Tc_{1,1}\right) ,  \label{b.4} 
\end{equation} 
\begin{equation} 
\left[ -\zeta ^{(0)}\left( T\partial _{T}+p\partial _{p}\right) -2\zeta 
^{(0)}+\nu \right] n_{1}T_{1}b_{1,1}=\frac{1}{3}\int d{\bf V}_{1}m_{1}{\bf V}
_{1}\cdot {\bf B}_{1}+\frac{T\zeta ^{(0)}}{p}n_{1}T_{1}c_{1,1},  \label{b.5} 
\end{equation} 
\begin{equation} 
\left[-\zeta ^{(0)}\left( T\partial _{T}+p\partial _{p}\right) -\frac{1}{2}
\zeta ^{(0)}+\nu \right] n_{1}T_{1}c_{1,1}=\frac{1}{3}\int d{\bf V}_{1}m_{1}
{\bf V}_{1}\cdot {\bf C}_{1}-\frac{p\zeta ^{(0)}}{2T}n_{1}T_{1}b_{1,1}. 
\label{b.6} 
\end{equation} 
Here, $\nu $ is a collision frequency defined by  
\begin{eqnarray} 
\nu  &=&\frac{1}{3n_{1}T_{1}}\int d{\bf V}_{1}m_{1}{\bf V}_{1}\cdot \left[  
{\cal L}_{1}f_{1,M}{\bf V}_{1}-\delta \gamma {\cal M}_{1}f_{2,M}{\bf V}_{2}
\right]   \nonumber \\ 
&=&-\frac{1}{3n_{1}T_{1}}\int d{\bf V}_{1}m_{1}{\bf V}_{1}\cdot \left( 
J_{12}[{\bf v}_{1}|f_{1,M}{\bf V}_{1},f_{2}^{(0)}]-\delta \gamma J_{12}[{\bf
v}_{1}|f_{1}^{(0)},f_{2,M}{\bf V}_{2}]\right) .  \label{b.7} 
\end{eqnarray} 
The self-collision terms of ${\cal L}_{i}$ arising from $J_{11}$ do not 
occur in Eq.\ (\ref{b.7}) since these conserve momentum for species $1$. The 
velocity integrals appearing in Eqs.\ (\ref{b.4})--(\ref{b.6}) can be 
performed using Eqs.\ (\ref{3.21})--(\ref{3.23})  
\begin{equation} 
\frac{1}{3}\int d{\bf V}_1m_{1}{\bf V}_{1}\cdot {\bf A}_{1}=-\left( \frac{
\partial }{\partial x_{1}}n_{1}T_{1}\right) _{p,T},  \label{b.8} 
\end{equation} 
\begin{equation} 
\frac{1}{3}\int d{\bf V}_{1}m_{1}{\bf V}_{1}\cdot {\bf B}_{1}=-\frac{
n_{1}T_{1}}{p}\left( 1-\frac{m_{1}nT}{\rho T_{1}}\right) ,  \label{b.9} 
\end{equation} 
\begin{equation} 
\frac{1}{3}\int d{\bf V}_{1}m_{1}{\bf V}_{1}\cdot {\bf C}_{1}=0. 
\label{b.10} 
\end{equation} 
From dimensional analysis $n_{1}T_{1}a_{1,1}\sim T^{1/2}$, $
n_{1}T_{1}b_{1,1}\sim T^{1/2}/p$, and $n_{1}T_{1}c_{1,1}\sim T^{-1/2}$ so the 
temperature derivatives can be performed in Eqs.\ (\ref{b.4})--(\ref{b.6}) 
and the result is  
\begin{equation} 
\left( \nu -\frac{1}{2}\zeta ^{(0)}\right) a_{1,1}=-\left( \frac{\partial }{
\partial x_{1}}\ln n_{1}T_{1}\right) _{p,T}+\left( \frac{\partial \zeta 
^{(0)}}{\partial x_{1}}\right) _{p,T}\left( pb_{1,1}+Tc_{1,1}\right) , 
\label{b.11} 
\end{equation} 
\begin{equation} 
\left( \nu -\frac{3}{2}\zeta ^{(0)}\right) b_{1,1}=-\frac{1}{p}\left( 1-\frac{
m_{1}nT}{\rho T_{1}}\right) +\frac{T\zeta ^{(0)}}{p}c_{1,1},  \label{b.12} 
\end{equation} 
\begin{equation} 
c_{1,1}=-\frac{p\zeta ^{(0)}}{2T\nu }b_{1,1}.  \label{b.13} 
\end{equation} 
The solutions are given by Eqs.\ (\ref{4.6})--(\ref{4.8}) of the text. 
 
The analysis for the coefficients  $\{a_{1,2}; b_{1,2}; c_{1,2}\}$ 
(Dufour coefficient, pressure energy coefficient, 
and thermal conductivity) is similar to that above for the mass flux.
The result is 
\begin{eqnarray}
\label{b.14}
\left[-\zeta^{(0)}\left(T\partial_{T}+p\partial 
_{p}\right)+\nu_{11}\right] 
T^3a_{1,2}&+&T^3\nu_{12}a_{2,2}=\frac{2}{15}\frac{m_1T^3}{n_1T_1^3}\int
d{\bf v}_1 {\bf S}_1\cdot {\bf A}_{12}\nonumber\\
& & 
+T^3\left( \frac{\partial \zeta ^{(0)}}
{\partial x_{1}}\right)_{p,T}(pb_{1,2}+Tc_{1,2}),
\end{eqnarray}
\begin{eqnarray}
\label{b.15}
\left[-\zeta^{(0)}\left(T\partial_{T}+p\partial 
_{p}\right)-2\zeta^{(0)}+\nu_{11}\right] 
T^3b_{1,2}&+&T^3\nu_{12}b_{2,2}=\frac{2}{15}\frac{m_1T^3}{n_1T_1^3}\int
d{\bf v}_1 {\bf S}_1\cdot {\bf B}_{12}\nonumber\\
& & 
+\frac{T^4 \zeta^{(0)}}{p}c_{1,2},
\end{eqnarray}
\begin{eqnarray}
\label{b.16}
\left[-\zeta^{(0)}\left(T\partial_{T}+p\partial 
_{p}\right)-\frac{1}{2}\zeta^{(0)}+\nu_{11}\right] 
T^3c_{1,2}&+&T^3\nu_{12}c_{2,2}=\frac{2}{15}\frac{m_1T^3}{n_1T_1^3}\int
d{\bf v}_1 {\bf S}_1\cdot {\bf C}_{12}\nonumber\\
& & 
-\frac{1}{2}pT^2\zeta^{(0)}c_{1,2}.
\end{eqnarray}
Here, the collision frequencies $\nu_{ii}$ and $\nu_{ij}$ are defined by Eqs.\  (\ref{4.3:13a}), respectively and 
(\ref{4.3:13b}) and  
 \begin{eqnarray}
\label{b.17}
{\bf A}_{12}&=& {\bf 
A}_1-\frac{m_1m_2n}{\rho 
n_1\gamma_1}\zeta^{(0)}\left(T\partial_{T}+p\partial 
_{p}\right)\frac{D}{T}{\bf V}_1f_{1,M}+\frac{m_1m_2n}{\rho n_1T_1}D
\nonumber\\
& & \times \left[{\cal L}_{1}(f_{1,M}{\bf V}_{1})-\delta\gamma{\cal 
M}_{1}(f_{2,M}{\bf V}_{1})\right]-\left( \frac{\partial \zeta ^{(0)}}
{\partial x_{1}}\right)_{p,T}\frac{\rho}{n_1T_1}\left(D_p+D'\right)
f_{1,M}{\bf V}_{1},
\end{eqnarray}
\begin{eqnarray}
\label{b.18}
{\bf B}_{12}&=& {\bf B}_1-\frac{\rho}{ 
n_1\gamma_1}\zeta^{(0)}\left(T\partial_{T}+p\partial 
_{p}\right)\frac{D_p}{pT}{\bf V}_1f_{1,M}+\frac{\rho}{pn_1T_1}D_p
\nonumber\\
& & \times \left[{\cal L}_{1}(f_{1,M}{\bf V}_{1})-\delta\gamma{\cal 
M}_{1}(f_{2,M}{\bf V}_{1})\right]-\zeta^{(0)}
\frac{\rho}{pn_1T_1}\left(D'-2D_p\right)f_{1,M}{\bf V}_{1},
\end{eqnarray}
\begin{eqnarray}
\label{b.19}
{\bf C}_{12}&=& {\bf C}_1-\frac{\rho}{ 
n_1\gamma_1}\zeta^{(0)}\left(T\partial_{T}+p\partial 
_{p}\right)\frac{D'}{T^2}{\bf V}_1f_{1,M}+\frac{\rho}{Tn_1T_1}D'
\nonumber\\
& & \times \left[{\cal L}_{1}(f_{1,M}{\bf V}_{1})-\delta\gamma{\cal 
M}_{1}(f_{2,M}{\bf V}_{1})\right]-\frac{1}{2}\zeta^{(0)}
\frac{\rho}{Tn_1T_1}\left(D'-D_p\right)f_{1,M}{\bf V}_{1},
\end{eqnarray}
where $\gamma_i=T_i/T$. The corresponding integral equations for $a_{2,2}$, 
$b_{2,2}$, and $c_{2,2}$ can be obtained from Eqs.\  (\ref{b.14})--(\ref{b.16}) by 
just making the change $1\leftrightarrow 2$. Dimensional analysis requires that 
$a_{1,2}\sim T^{-3/2}$, $b_{1,2}\sim T^{-3/2}/p$, and $a_{1,2}\sim T^{-5/2}$. When one 
takes into account this dependence and the forms of ${\bf A}_{ij}$,  ${\bf B}_{ij}$, and
${\bf C}_{ij}$, one finally arrives to Eq.\  (\ref{4.3:12}).

\section{Evaluation of $\protect\nu $} 
\label{appB} 
 
The collision frequency $\nu $ is defined by the collision integrals in Eq.\ 
(\ref{b.7}). To simplify the integrals, a useful identity for an arbitrary 
function $h({\bf v}_{1})$ is given by  
\begin{eqnarray} 
\int d{\bf v}_{1}h({\bf v}_{1})J_{ij}\left[ {\bf v}_{1}|f_{i},f_{j}\right] 
&=&\sigma _{ij}^{2}\int \,d{\bf v}_{1}\,\int \,d{\bf v}_{2}f_{i}({\bf r},  
{\bf v}_{1},t)f_{j}({\bf r},{\bf v}_{2},t)  \nonumber \\ 
&&\times \int d\widehat{\bbox {\sigma }}\,\Theta (\widehat{\bbox {\sigma}} 
\cdot {\bf g}_{12})(\widehat{\bbox {\sigma }}\cdot {\bf g}_{12})\,\left[ h(  
{\bf v}_{1}^{^{\prime \prime }})-h({\bf v}_{1})\right] \;,  \label{a1} 
\end{eqnarray} 
with  
\begin{equation} 
{\bf v}_{1}^{^{\prime \prime }}={\bf v}_{1}-\mu _{ji}(1+\alpha _{ij})(  
\widehat{\bbox {\sigma }}\cdot {\bf g}_{12})\widehat{\bbox {\sigma}}\;. 
\label{a2} 
\end{equation} 
This result applies for both $i=j$ and $i\neq j$. Use of this in Eq.\ (\ref {b.7}) gives  
\begin{eqnarray} 
\nu &=&-\frac{m_{1}}{3n_{1}T_{1}}\int d{\bf V}_{1}{\bf V}_{1}\cdot \left( 
J_{12}[f_{1,M}{\bf V}_{1},f_{2}^{(0)}]-\delta \gamma 
J_{12}[f_{1}^{(0)},f_{2,M}{\bf V}_{2}]\right)  \nonumber \\ 
&=&\frac{1}{6}\pi \sigma _{12}^{2}\mu _{21}(1+\alpha _{12})\frac{m_{1}}{ 
n_{1}T_{1}}\int \,d{\bf V}_{1}\,\int \,d{\bf V}_{2}\,g_{12}\left[ f_{1,M}(  
{\bf V}_{1})f_{2}^{(0)}({\bf V}_{2})({\bf V}_{1}\cdot {\bf g}_{12})\right.  
\nonumber \\ 
&&\left. -\delta \gamma f_{1}^{(0)}({\bf V}_{1})f_{2,M}({\bf V}_{2})({\bf V} 
_{2}\cdot {\bf g}_{12})\right],  \label{a.3} 
\end{eqnarray} 
where $\gamma=T_1/T_2$ and $\delta=n_1/n_2$. Substitution of the 
distribution functions from Eqs.\ (\ref{4.02}) and (\ref{4.1}) gives  
\begin{eqnarray} 
\nu &=&\frac{n_{2}}{3\pi ^{2}\left( 1+\mu \right) }(1+\alpha _{12})\sigma 
_{12}^{2}\theta _{1}^{5/2}\theta _{2}^{3/2}v_{0}\int \,d{\bf V}_{1}^{\ast 
}\,\int \,d{\bf V}_{2}^{\ast }e^{-\theta _{1}V_{1}^{\ast 2}}e^{-\theta 
_{2}V_{2}^{\ast 2}}g_{12}^{\ast }\left\{ \left[ 1+\frac{c_{2}}{4}\right. 
\right.  \nonumber \\ 
&\times &\left. \left. \left( \theta _{2}^{2}V_{2}^{\ast 4}-5\theta 
_{2}V_{2}^{\ast 2}+\frac{15}{4}\right) \right] ({\bf V}_{1}^{\ast}\cdot  
{\bf g}_{12}^{\ast})-\delta \gamma \left[ 1+\frac{c_{1}}{4}\left( \theta 
_{1}^{2}V_{1}^{\ast 4}-5\theta _{1}V_{1}^{\ast 2}+\frac{15}{4}\right) 
\right] ({\bf V}_{2}^{\ast}\cdot {\bf g}_{12}^{\ast })\right\}  \nonumber 
\\ &=&\frac{(1+\alpha _{12})\sigma _{12}^{2}}{3\pi ^{2}\gamma _{1}\gamma 
_{2}} \left( \theta _{1}\theta _{2}\right) ^{3/2}v_{0}\left[ n_{2}\gamma 
_{2}X\left( c_{2},\theta _{1},\theta _{2}\right) +n_{1}\gamma _{1}X\left( 
c_{1},\theta _{2},\theta _{1}\right) \right].  \label{a.4} 
\end{eqnarray} 
Here, $\mu=m_1/m_2$, ${\bf V}_{i}^{\ast }={\bf V}_{i}/v_{0}$, ${\bf g}
_{12}^{\ast }={\bf g}_{12}/v_{0}$, $\theta _{i}=(\mu _{ji}\gamma _{i})^{-1}$, and $v_{0}=\sqrt{2T(m_{1}+m_{2})/m_{1}m_{2}}$. In 
addition, the quantities $X\left(c_{2},\theta _{1},\theta _{2}\right)$ and 
$I\left( \theta _{1},\theta_{2}\right)$ are given by  
\begin{equation} 
X\left( c_{2},\theta _{1},\theta _{2}\right) =\left[ 1+\frac{c_{2}}{4}\left( 
\theta _{2}^{2}\frac{d^{2}}{d\theta _{2}^{2}}+5\theta _{2}\frac{d}{d\theta 
_{2}}+\frac{15}{4}\right) \right] I\left( \theta _{1},\theta _{2}\right) , 
\label{a.5} 
\end{equation} 
\begin{equation} 
I\left( \theta _{1},\theta _{2}\right) =\int \,d{\bf V}_{1}^{\ast}\,\int 
\,d {\bf V}_{2}^{\ast }e^{-\theta _{1}V_{1}^{\ast 2}}e^{-\theta 
_{2}V_{2}^{\ast 2}}g_{12}^{\ast}({\bf V}_{1}^{\ast}\cdot {\bf g}
_{12}^{\ast}).  \label{a.6} 
\end{equation} 
The integral $I\left(\theta_{1},\theta_{2}\right)$ can be performed by the 
change of variables  
\begin{equation} 
{\bf x}={\bf V}_{1}^{\ast}-{\bf V}_{2}^{\ast},\quad {\bf y}=\theta _{1}  
{\bf V}_{1}^{\ast}+\theta _{2}{\bf V}_{2}^{\ast},  \label{a.7} 
\end{equation} 
with the Jacobian $\left( \theta _{1}+\theta _{2}\right) ^{-3}$. The 
integral becomes  
\begin{equation} 
I\left( \theta _{1},\theta _{2}\right) =4\pi ^{5/2}\frac{\left( \theta 
_{1}+\theta _{2}\right) ^{1/2}}{\theta _{2}^{2}\theta _{1}^{3}}.  \label{a.8} 
\end{equation} 
Use of this result in (\ref{a.5}) and (\ref{a.3}) gives the desired result  
\begin{eqnarray} 
\nu &=&\frac{4}{3}\sqrt{\pi }\sigma _{12}^{2}v_{0}\frac{1+\alpha _{12}}{ 
\gamma _{1}\gamma _{2}}\left( \theta _{1}\theta _{2}(\theta _{1}+\theta 
_{2})\right) ^{-3/2}\left[ \left( n_{2}\gamma _{2}\theta _{2}+n_{1}\gamma 
_{1}\theta _{1}\right) \left( \theta _{1}+\theta _{2}\right) ^{2}\right.  
\nonumber \\ 
&&\left. -\frac{1}{16}\left( c_{1}n_{1}\gamma _{1}\theta _{1}\theta 
_{2}^{2}+c_{2}n_{2}\gamma _{2}\theta _{2}\theta _{1}^{2}\right) \right] . 
\label{a.9} 
\end{eqnarray} 
This leads directly to the result given by Eq.\ (\ref{4.11}) in the text.

\section{Evaluation of \lowercase{$\tau_{ii}$} and 
\lowercase{$\tau_{ij}$} }
\label{appC} 

The collision frequencies $\tau_{ii}$ and $\tau_{ij}$ appearing in the 
evaluation of the shear viscosity are defined by Eqs.\ (\ref{4.2:6}) and 
(\ref{4.2:7}).  All of these have the form 
\begin{eqnarray}
\label{c1}
\int d{\bf v}_1 
{\bf V}_{1}{\bf V}_{1} J_{ij}[{\bf v}_1|f_i,f_j]&=&\sigma_{ij}^2\int\, d{\bf 
v}_1\int\, d{\bf v}_2\, f_i({\bf v}_1) f_j({\bf v}_2)\nonumber\\
& & \times \int d\widehat{\bbox {\sigma}}\,\Theta (\widehat{\bbox {\sigma}}
\cdot {\bf g}_{12})(\widehat{\bbox {\sigma}}\cdot {\bf g}_{12})\,\left[
{\bf V}_1''{\bf V}_{1}''-{\bf V}_{1}{\bf V}_{1}\right],
\end{eqnarray}
where the identity (\ref{a1}) has been used. 
This result applies for both $i=j$ and $i\neq j$. Using the scattering rule (\ref{a2}), the last term on the right hand side can be explicitly computed as
\begin{eqnarray}
\label{c2}
{\bf V}_{1}''{\bf V}_{1}''-{\bf V}_{1}{\bf V}_{1}&=&-\mu_{ji}(1+
\alpha_{ij})(\widehat{\bbox {\sigma}}\cdot {\bf g}_{12})\left[
{\bf G}_{ij}{\bbox{\sigma}}+{\bbox{\sigma}}{\bf G}_{ij}\right.\nonumber\\
& &
\left.+\mu_{ji}({\bf g}_{12}{\bbox{\sigma}}+{\bbox{\sigma}}{\bf g}_{12}) 
-\mu_{ji}(1+\alpha_{ij})(\widehat{\bbox {\sigma}}\cdot {\bf g}_{12})
{\bbox{\sigma}}{\bbox{\sigma}}\right]
\end{eqnarray}
Here, ${\bf G}_{ij}=\mu_{ij}{\bf V}_1+\mu_{ji}{\bf V}_2$. 
Substitution of (\ref{c2}) into Eq.\ 
(\ref{c1}) allows the angular integral to be performed with the result
\begin{eqnarray}
\label{c3}
\int d\widehat{\bbox {\sigma}}\,\Theta (\widehat{\bbox {\sigma}}\cdot {\bf g
}_{12})(\widehat{\bbox {\sigma}}\cdot {\bf g}_{12})\, &&\left[ 
{\bf V}_{1}''{\bf V}_{1}''-{\bf V}_{1}{\bf V}_{1}\right]= 
-\frac{m_i}{2}\pi \mu_{ji}(1+\alpha _{ij})\left[g_{12}({\bf G}_{ij}{\bf g}_{12}+
{\bf g}_{12}{\bf G}_{ij})\right. \nonumber\\
& &\left.
+\frac{\mu_{ji}}{2}
(3-\alpha_{ij})g_{12}{\bf g}_{12}{\bf g}_{12}-\frac{\mu_{ji}}{6}
(1+\alpha_{ij})g^3\openone\right].
\end{eqnarray}
Notice that the last term in (\ref{c3}) vanishes when is contracted with the traceless tensor ${\sf R}_{i}$. Now, the different collision integrals can be easily calculated by the 
same method as described in Appendix \ref{appB}. After a lengthy calculation, one gets
\begin{eqnarray}
\label{c4}
{\cal A}_{12}&=&\int d{\bf v}_1 
R_{1,\alpha\beta}J_{12}[f_1^{(0)}, f_{2,M}R_{2,\alpha\beta}]\nonumber\\
&=&
-\frac{4}{3}m_1m_2n_1n_2\sqrt{\pi}\mu_{21}(1+\alpha_{12})\sigma_{12}^2 
v_0^5 \left(\theta_1\theta_2\right)^{-1/2}\nonumber\\
& & \times 
\left[6 \theta_2^{-2}(\mu_{12}\theta_2-\mu_{21}\theta_1)
\left(\theta_1+\theta_2\right)^{-1/2}+\case{3}{2}\mu_{21}
\theta_2^{-2}\left(\theta_1+\theta_2\right)^{1/2}(3-\alpha_{12})-
5\theta_2^{-1}\left(\theta_1+\theta_2\right)^{-1/2} \right.\nonumber\\
& &\left. +\frac{c_1}{16}\frac{2\theta_1(10-12\mu_{12}-9\mu_{21})+\theta_2
(5-6\mu_{12})-\case{3}{2}\mu_{21}(3-\alpha_{12})\left(\theta_1+\theta_2\right)}
{\left(\theta_1+\theta_2\right)^{5/2}}\right],
\end{eqnarray} 
\begin{eqnarray}
\label{c5}
{\cal B}_{12}&=&
\int d{\bf v}_1 
R_{1,\alpha\beta}J_{12}[f_{1,M}R_{1,\alpha\beta},f_2^{(0)}]\nonumber\\
&=&
-\frac{4}{3}m_1^2n_1n_2\sqrt{\pi}\mu_{21}(1+\alpha_{12})\sigma_{12}^2 
v_0^5 \left(\theta_1\theta_2\right)^{-1/2}\nonumber\\
& & \times 
\left[6 \theta_1^{-2}(\mu_{12}\theta_2-\mu_{21}\theta_1)
\left(\theta_1+\theta_2\right)^{-1/2}+\case{3}{2}\mu_{21}\theta_1^{-2}
\left(\theta_1+\theta_2\right)^{1/2}(3-\alpha_{12})+
5\theta_1^{-1}\left(\theta_1+\theta_2\right)^{-1/2} \right.\nonumber\\
& &\left. +\frac{c_2}{16}\frac{2\theta_2(12\mu_{21}+9\mu_{12}-10)-\theta_1
(5-6\mu_{21})-\case{3}{2}\mu_{21}(3-\alpha_{12})
\left(\theta_1+\theta_2\right)}
{\left(\theta_1+\theta_2\right)^{5/2}}\right].
\end{eqnarray}
\begin{eqnarray}
\label{c6}
{\cal A}_{11}+{\cal B}_{11}&=&
\int d{\bf v}_1 
R_{1,\alpha\beta}\left\{J_{11}[f_1^{(0)},f_{1,M}R_{1,\alpha\beta}]
+J_{11}[f_{1,M}R_{1,\alpha\beta},f_1^{(0)}]\right\}\nonumber\\
&=&-32m_1^2
n_1^2\sqrt{\pi}(1+\alpha_{11})\sigma_1^2 (T_1/m_1)^{5/2}
\left[1-\frac{1}{4}
(1-\alpha_{11})^2\right](1-\frac{c_1}{64}).
\end{eqnarray}
In the case of mechanically equivalent particles ($m_1=m_2=m$, 
$\sigma_{ij}=\sigma$, $\alpha_{ij}=\alpha$, $c_i=c$), Eqs.\ 
(\ref{c4})--(\ref{c6}) reduce to those previously calculated 
in the single gas case in the determination of the shear viscosity. \cite{BDKS98}

This completely determines $\tau_{11}$ and $\tau_{12}$. The corresponding 
expressions for $\tau_{22}$ and $\tau_{21}$ can be inferred from Eqs.\ (\ref{c4}), (\ref{c5}), and (\ref{c6}) by interchanging 
$1\leftrightarrow 2$.  

\section{Evaluation of $Y$\lowercase{$_{i}$}, 
\lowercase{$\nu_{ii}$} and \lowercase{$\nu_{ij}$}} 
\label{appD}

The collision frequencies $\nu_{ii}$, $\nu_{ij}$ and $Y_i$ $(i=1,\ldots 6)$ 
that determine the coefficients $a_{i,2}$, $b_{i,2}$ and $c_{i,2}$ for the heat
flux are defined by the collision integrals (\ref{4.3:13a}), (\ref{4.3:13b}), 
(\ref{4.3:14}), (\ref{4.3:15}), (\ref{4.3:16}), and (\ref{4.3:17}). All of these have the form 
\begin{eqnarray}
\label{d1}
\int d{\bf v}_1 
{\bf S}_{i}({\bf V}_1) J_{ij}[{\bf v}_1|f_i,f_j]&=&\sigma_{ij}^2\int\, d{\bf 
v}_1\int\, d{\bf v}_2\, f_i({\bf v}_1) f_j({\bf v}_2)\nonumber\\
& & \times \int d\widehat{\bbox {\sigma}}\,\Theta (\widehat{\bbox {\sigma}}
\cdot {\bf g}_{12})(\widehat{\bbox {\sigma}}\cdot {\bf g}_{12})\,\left[
{\bf S}_i({\bf V}_{1}'')-{\bf S}_i({\bf V}_{1})\right].
\end{eqnarray}
Using the scattering rule (\ref{a2}), the last term on the right hand 
side can be explicitly computed as
\begin{eqnarray}
\label{d2}
{\bf S}_{i}({\bf V}_{1}^{^{\prime \prime }})-{\bf S}_{i}({\bf V}_{1}) &=&
\frac{m_{i}}{2}(1+\alpha _{ij})\mu_{ji}(\widehat{\bbox {\sigma}}\cdot {\bf g
}_{12})\left\{ \left[ (1-\alpha_{ij}^{2})\mu_{ji}^{2}(\widehat{\bbox
{\sigma}}\cdot {\bf g}_{12})^{2}-G_{ij}^{2}-\mu_{ji}^{2}g_{12}^{2}\right.
\right.   \nonumber \\
&&\left. -2\mu_{ji}({\bf g}_{12}\cdot {\bf G}_{ij})+2(1+\alpha_{ij})\mu
_{ji}(\widehat{\bbox {\sigma}}\cdot {\bf g}_{12})(\widehat{\bbox {\sigma }}
\cdot {\bf G}_{ij})-\frac{5T_{i}}{m_{i}}\right] \widehat{\bbox {\sigma}} 
\nonumber \\
&&-\left[ (1-\alpha_{ij})\mu_{ji}(\widehat{\bbox {\sigma }}\cdot {\bf g}_{12}
)+2(\widehat{\bbox {\sigma}}\cdot {\bf G}_{ij})\right] {\bf G}_{ij}
\nonumber \\
&&\left. -\mu_{ji}\left[(1-\alpha_{ij})\mu_{ji}(\widehat{\bbox {\sigma}}
\cdot {\bf g}_{12})+2(\widehat{\bbox{\sigma}}\cdot {\bf G}_{ij})
\right] {\bf g}_{12}\right\} \;. 
\end{eqnarray}
Substitution of (\ref{d2}) into (\ref{d1}) allows the angular integral to be performed with 
the result
\begin{eqnarray}
\label{d3}
\int d\widehat{\bbox {\sigma}}\,\Theta (\widehat{\bbox {\sigma}}\cdot {\bf g
}_{12})(\widehat{\bbox {\sigma}}\cdot {\bf g}_{12})\, &&\left[ {\bf S}_{i}(
{\bf V}_{1}^{^{\prime \prime }})-{\bf S}_{i}({\bf V}_{1})\right] =-\frac{
m_{i}}{2}\pi (1+\alpha_{ij})\mu_{ji}\left\{ \left[ \frac{1}{2}
g_{12}G_{ij}^{2}\right. \right.   \nonumber \\
&&\left. +\frac{1}{6}\mu_{ji}^{2}\left( 2\alpha_{ij}^{2}-3\alpha
_{ij}+4\right) g_{12}^{3}-\frac{1}{2}\mu_{ji}\left( \alpha_{ij}-3\right)
g_{12}\left( {\bf g}_{12}\cdot {\bf G}_{ij}\right) -\frac{5T_{i}}{2m_{i}}
g_{12}\right] {\bf g}_{12}  \nonumber \\
&&\left. +\left[ g_{12}\left( {\bf g}_{12}\cdot {\bf G}_{ij}\right) 
-\frac{1}{3}\mu_{ji}\left( 2\alpha_{ij}-1\right) g_{12}^{3}\right] 
{\bf G}_{ij}\right\} \;.  
\end{eqnarray}
Now the different collision integrals can be evaluated. 

\subsection{Evaluation of $Y_{i}$}

The coefficients $Y_{1,3,5}$  are obtained from Eqs.\ 
(\ref{4.3:15})--(\ref{4.3:17}).  
The collision integrals appearing in these expressions can be evaluated 
directly by using identical mathematical steps as before. After some algebra, 
the result is   
\begin{eqnarray}
\label{d4}
{\cal C}_{12}&=&\int d{\bf v}_1 
{\bf S}_{1}\cdot J_{12}[f_1^{(0)},f_{2,M}{\bf V}_2] \nonumber\\
&=&
-\frac{1}{2}m_1n_1n_2\sqrt{\pi}\mu_{21}(1+\alpha_{12})\sigma_{12}^2 
v_0^5\left(\theta_1+\theta_2\right)^{-1/2}\left(\theta_1\theta_2
\right)^{-3/2}\nonumber\\
& & \times 
\left\{5(2\beta_{12}-\theta_1)+\mu_{21}(\theta_1+\theta_2)\left[5(1-\alpha_{12})+2
(7\alpha_{12}-11)\beta_{12}\theta_2^{-1}\right]\right.\nonumber\\
& &\left. 
-18\beta_{12}^2\theta_2^{-1}-2\mu_{21}^2\left(2\alpha_{12}^{2}-3\alpha
_{12}+4\right)\theta_2^{-1}(\theta_1+\theta_2)^2+
5(\theta_1+\theta_2)\right]\nonumber\\
& & -\frac{c_1}{16}\left(\theta_1+\theta_2\right)^{-2}\theta_2\left\{
4\theta_1^2\left[-5+36\mu_{12}^2+5\mu_{21}(\alpha_{12}-3)
\right.\right.\nonumber\\
& & \left.
+\mu_{21}^2(25-8\alpha_{12}+3\alpha_{12}^2)-2\mu_{12}(10+\mu_{21}(
7\alpha_{12}-29))\right]\nonumber\\
& & 
+\theta_2^2\left[5+54\mu_{12}^2+15\mu_{21}(\alpha_{12}-1)
+6\mu_{21}^2(4-3\alpha_{12}+2\alpha_{12}^2)\right.\nonumber\\
& & 
\left. -6\mu_{12}(5+\mu_{21}(7\alpha_{12}-11))\right]-\theta_1\theta_2\left[
-144\mu_{12}^2+2\mu_{12}(40+\mu_{21}(49\alpha_{12}-95))
\right.\nonumber\\
& & \left.\left.\left.
+\mu_{21}(45-35\alpha_{12}-2\mu_{21}(35-25\alpha_{12}+
12\alpha_{12}^2)\right]\right\}
\right\},
\end{eqnarray} 
\begin{eqnarray}
\label{d5}
{\cal D}_{12}&=&\int d{\bf v}_1 
{\bf S}_{1}\cdot J_{12}[f_{1,M}{\bf V}_1,f_2^{(0)}]\nonumber\\
&=&-\frac{1}{2}m_1n_1n_2\sqrt{\pi}\mu_{21}(1+\alpha_{12})\sigma_{12}^2 
v_0^5\left(\theta_1+\theta_2\right)^{-1/2}\left(\theta_1\theta_2
\right)^{-3/2}\nonumber\\
& & \times 
\left\{5(2\beta_{12}+\theta_2)+\mu_{21}(\theta_1+\theta_2)
\left[5(1-\alpha_{12})-2
(7\alpha_{12}-11)\beta_{12}\theta_1^{-1}\right]\right.\nonumber\\
& &\left. 
+18\beta_{12}^2\theta_1^{-1}+2\mu_{21}^2\left(2\alpha_{12}^{2}-3\alpha
_{12}+4\right)\theta_1^{-1}(\theta_1+\theta_2)^2-
5\theta_2\theta_1^{-1}(\theta_1+\theta_2)\right]\nonumber\\
& & +\frac{c_2}{16}\left(\theta_1+\theta_2\right)^{-2}\theta_1\left\{
3\theta_1^2\mu_{21}(1+\alpha_{12})\left[4\mu_{21}(1+\alpha_{12})-5\right]
\right.\nonumber\\
& &
+\theta_2^2\left[-15+54\mu_{12}^2-20\mu_{21}(3+\alpha_{12})+2\mu_{21}^2(
40+19\alpha_{12}+6\alpha_{12}^2)\right.\nonumber\\
& & \left.
+2\mu_{12}(-20+\mu_{21}(61+7\alpha_{12}))\right]
+\theta_1\theta_2\left[2\mu_{12}(-5+7\mu_{21}(1+\alpha_{12}))\right.
\nonumber\\
& &\left.\left.\left. 
+\mu_{21}(-5(9+7\alpha_{12})+\mu_{21}(38+62\alpha_{12}+
24\alpha_{12}^2)\right]\right\}\right\},
\end{eqnarray}
\begin{eqnarray}
\label{d6}
{\cal C}_{11}+{\cal D}_{11}&=&
\int d{\bf v}_1 {\bf S}_{1}\cdot \left\{J_{11}[f_1^{(0)},f_{1,M}{\bf V}_2]
+J_{11}[f_{1,M}{\bf 
V}_1,f_1^{(0)}]\right\}\nonumber\\
&=&
-10\sqrt{\pi}m_1\sigma_{11}^2n_1^2(T_1/m_1)^{5/2}
(1+\alpha_{11})\left[1-\alpha_{11}+\frac{1}{320}c_1(21\alpha_{11}-53)\right].
\end{eqnarray}
In the above expressions we have introduced the quantity
\begin{equation}
\label{d7}
\beta_{12}=\mu_{12}\theta_2-\mu_{21}\theta_1.
\end{equation}
This completely determines the coefficients $Y_{1,3,5}$. The corresponding 
expressions for $Y_{2,4,6}$ can be inferred from 
Eqs.\  (\ref{d5})--(\ref{d7}) by interchanging  $1\leftrightarrow 2$.

\subsection{Evaluation of $\nu_{ii}$ and $\nu_{ij}$}

The collision frequencies $\nu_{ii}$ and $\nu_{ij}$ are defined
by Eqs.\ (\ref{4.3:13a}) and (\ref{4.3:13b}).
These collision integrals are evaluated in the same manner as
those for $Y_i$. The result is
\begin{equation}
\label{d9}
\int d{\bf v}_1 
{\bf S}_{1}\cdot J_{12}[f_1^{(0)},f_{2,M}{\bf S}_2]=
\int d{\bf v}_1 
{\bf S}_{1}\cdot J_{12}[f_1^{(0)},f_{2,M}\frac{m_2}{2}V_2^2{\bf V}_2
-\case{5}{2}T_2{\bf V}_2]={\cal F}_{12}-\frac{5}{2}T_2{\cal C}_{12},
\end{equation}
\begin{eqnarray}
\label{d10}
{\cal F}_{12}&=&\int d{\bf v}_1 
{\bf S}_{1}\cdot J_{12}[f_1^{(0)},f_{2,M}\frac{m_2}{2}V_2^2{\bf V}_2]\nonumber\\
&=& 
\frac{1}{8}m_1n_1n_2\sqrt{\pi}\mu_{21}(1+\alpha_{12})\sigma_{12}^2 
v_0^7\left(\theta_1+\theta_2\right)^{-3/2}\left(\theta_1\theta_2
\right)^{-3/2}\nonumber\\
& & \times 
\left\{
2\mu_{21}^2\theta_2^{-2}(\theta_1+\theta_2)^2
\left(2\alpha_{12}^{2}-3\alpha_{12}+4\right)
(8\theta_1+5\theta_2)\right.\nonumber\\
& & 
-\mu_{21}(\theta_1+\theta_2)
\left[2\beta_{12}\theta_2^{-2}(8\theta_1+5\theta_2)(7\alpha_{12}
-11)-2\theta_1\theta_2^{-1}(29\alpha_{12}-37)+25(1-\alpha_{12})\right]
\nonumber\\
& & 
+18\beta_{12}^2\theta_2^{-2}(8\theta_1+5\theta_2)-
2\beta_{12}\theta_2^{-1}(66\theta_1+25\theta_2)
\nonumber\\
& & 
+5\theta_1\theta_2^{-1}
(6\theta_1+11\theta_2)-5(\theta_1+\theta_2)\theta_2^{-1}(6\theta_1+5\theta_2)
\nonumber\\
& & +\frac{c_1}{16}(\theta_1+\theta_2)^{-2}\left\{16\theta_1^3\left[
5+72\mu_{12}^2+\mu_{12}(-66+\mu_{21}(137-7\alpha_{12}))
\right.\right.\nonumber\\
& & \left.
-2\mu_{21}(34+\alpha_{12})
+\mu_{21}^2(68-\alpha_{12}+3\alpha_{12}^2)\right]\nonumber\\
& & 
+5\theta_2^3\left[5+54\mu_{12}^2-15\mu_{21}(1-\alpha_{12})+6\mu_{21}^2
(4-3\alpha_{12}+2\alpha_{12}^2)-6\mu_{12}(5+\mu_{21}(7\alpha_{12}-11)\right]
\nonumber\\
& & 
+2\theta_1^2\theta_2\left[-170+504\mu_{12}^2+\mu_{21}(55\alpha_{12}-17)
+2\mu_{21}^2(151-62\alpha_{12}+39\alpha_{12}^2)\right.\nonumber\\
& & \left.
-8\mu_{12}(9+7\mu_{21}
(5\alpha_{12}-13))\right]-\theta_1\theta_2^2
\left[20-936\mu_{12}^2+\mu_{21}(251-217\alpha_{12})
\right.\nonumber\\
& & \left.\left.
+\mu_{21}^2(-446+322\alpha_{12}-168\alpha_{12}^2)
+2\mu_{12}(234+\mu_{21}
(329\alpha_{12}-607))\right]\right\},
\end{eqnarray} 
\begin{equation}
\label{d11}
\int d{\bf v}_1 
{\bf S}_{1}\cdot J_{12}[f_{1,M}{\bf S}_1,f_2^{(0)}]=
\int d{\bf v}_1 
{\bf S}_{1}\cdot J_{12}[f_{1,M}\frac{m_1}{2}V_1^2{\bf V}_1
-\case{5}{2}T_1{\bf V}_1, f_2^{(0)}]=
{\cal H}_{12}-\frac{5}{2}T_1{\cal D}_{12},
\end{equation}
\begin{eqnarray}
\label{d12}
{\cal H}_{12}&=&\int d{\bf v}_1 
{\bf S}_{1}\cdot J_{12}[f_{1,M}\frac{m_1}{2}V_1^2{\bf V}_1,f_2^{(0)}]
\nonumber\\
&=& 
-\frac{1}{8}m_1n_1n_2\sqrt{\pi}\mu_{21}(1+\alpha_{12})\sigma_{12}^2 
v_0^7\left(\theta_1+\theta_2\right)^{-3/2}\left(\theta_1\theta_2
\right)^{-3/2}\nonumber\\
& & \times 
\left\{
2\mu_{21}^2\theta_1^{-2}(\theta_1+\theta_2)^2
\left(2\alpha_{12}^{2}-3\alpha_{12}+4\right)
(5\theta_1+8\theta_2)\right.\nonumber\\
& & 
-\mu_{21}(\theta_1+\theta_2)
\left[2\beta_{12}\theta_1^{-2}(5\theta_1+8\theta_2)(7\alpha_{12}
-11)+2\theta_2\theta_1^{-1}(29\alpha_{12}-37)-25(1-\alpha_{12})\right]
\nonumber\\
& & 
+18\beta_{12}^2\theta_1^{-2}(5\theta_1+8\theta_2)+
2\beta_{12}\theta_1^{-1}(66\theta_1+25\theta_2)
\nonumber\\
& & 
+5\theta_2\theta_1^{-1}
(11\theta_1+6\theta_2)-5(\theta_1+\theta_2)
\theta_1^{-2}\theta_2(6\theta_1+5\theta_2)
\nonumber\\
& & +\frac{c_2}{16}(\theta_1+\theta_2)^{-2}\left\{15\theta_1^3
\mu_{21}(1+\alpha_{12})(4\mu_{21}(1+\alpha_{12})-5)
\right.\nonumber\\
& & +2\theta_2^3\left[45+540\mu_{12}^2+16
\mu_{21}(\alpha_{12}-36)
+4\mu_{21}^2(134+5\alpha_{12}+6\alpha_{12}^2)\right.
\nonumber\\
& & \left.
-4\mu_{12}(148+\mu_{21}(7\alpha_{12}-263))\right]+
\theta_1^2\theta_2\left[-30-\mu_{21}(267+217\alpha_{12})
\right.\nonumber\\
& & \left.
+14\mu_{21}^2
(17+29\alpha_{12}+12\alpha_{12}^2)
+10\mu_{12}(7\mu_{21}(1+\alpha_{12}-5))\right]
\nonumber\\
& & 
+\theta_1\theta_2^2\left[-315+270\mu_{12}^2-2\mu_{21}(55\alpha_{12}+57)
+\mu_{21}^2(440+326\alpha_{12}+156\alpha_{12}^2)\right.\nonumber\\
& & \left.\left.
+2\mu_{12}(-2+\mu_{21}(7\alpha_{12}+277))\right]\right\},
\end{eqnarray}
\begin{eqnarray}
\label{d13}
\int d{\bf v}_1 {\bf S}_{1}\cdot \left\{J_{11}[f_1^{(0)},f_{1,M}{\bf S}_1]
+J_{11}[f_{1,M}{\bf S}_1,f_1^{(0)}]\right\}&=&
-8\sqrt{\pi}n_1^2\sigma_{12}^2m_1T_1(T_1/m_1)^{5/2}(1+\alpha_{11})
\nonumber\\
& \times & 
\left[1+\frac{33}{16}(1-\alpha_{11})+
\frac{1}{1024}c_1(19-3\alpha_{11})\right]. \nonumber\\
\end{eqnarray}
In these expressions, ${\cal C}_{12}$ and ${\cal D}_{12}$ are given by
Eqs.\ (\ref{d4}) and (\ref{d5}), respectively.

In the case of mechanically equivalent particles, the expression 
(\ref{d13}) coincides with the one previously obtained in the context of
determining the thermal conductivity in a one component
granular gas. \cite{BDKS98} From Eqs.\ (\ref{d10})--(\ref{d13}), one
easily gets the expressions for $\nu_{22}$ and $\nu_{21}$
by interchanging $1\leftrightarrow 2$.

\begin{figure}[tbp] 
\caption{Plot of the reduced pressure diffusion coefficient $D_p(\protect% 
\alpha)/D_p(1)$ as a function of the restitution coefficient $\protect\alpha% 
\equiv \protect\alpha_{11}= \protect\alpha_{22}=\protect\alpha_{12}$ for $% 
\protect\sigma_{11}= \protect\sigma_{22}=\protect\sigma_{12}$, a 
concentration ratio $n_1/n_2=0.25$, and two different values of the mass 
ratio: $m_1/m_2= 0.5$ and $m_1/m_2=4$. The dashed line refers to the case $% 
m_1/m_2=4$ by assuming the equality of the partial temperatures $\protect% 
\gamma=T_1/T_2=1$.} 
\label{fig1} 
\end{figure} 
 
\begin{figure}[tbp] 
\caption{Plot of the reduced diffusion coefficient $D(\protect\alpha)/D(1)$ 
as a function of the restitution coefficient $\protect\alpha\equiv \protect% 
\alpha_{11}= \protect\alpha_{22}=\protect\alpha_{12}$ for $\protect\sigma% 
_{11}= \protect\sigma_{22}=\protect\sigma_{12}$, a concentration ratio $% 
n_1/n_2=0.25$, and three different values of the mass ratio: $m_1/m_2= 0.5$,  
$m_1/m_2=1$, and $m_1/m_2=4$.} 
\label{fig2} 
\end{figure} 
 

\begin{references} 
\bibitem{BDKS98}  J. J. Brey, J. W. Dufty, C-S Kim, and A. Santos, 
``Hydrodynamics for granular flow at low density,'' Phys. Rev. E {\bf 58}, 
4638 (1998). 
 
\bibitem{JR85}  J. T. Jenkins and M. W. Richman, ``Kinetic theory for plane 
flows of a dense gas of identical, rough, inelastic, circular disks,'' Phys. 
Fluids {\bf 28}, 3485 (1986). 
 
\bibitem{LSJCh84}  C. K. W. Lun, S. B. Savage, D. J. Jeffrey, and N. 
Chepurniy, ``Kinetic theories for granular flow: inelastic particles in 
Couette flow and sligthly inelastic particles in a general flowfield,'' 
J. Fluid Mech. {\bf 140}, 223 (1984). 
 
\bibitem{GS95}  A. Goldshtein and M. Shapiro, ``Mechanics of collisional motion of granular materials. Part 1. General hydrodynamic equations,'' 
J. Fluid Mech. {\bf 282}, 75 (1995). 
 
 
\bibitem{GD199}  V. Garz\'{o} and J. W. Dufty, ``Dense fluid transport for inelastic hard spheres,'' Phys. Rev. E {\bf 59}, 5895 (1999). 
 
\bibitem{JM89}  J. T. Jenkins and F. Mancini, ``Kinetic theory for binary 
mixtures of smooth, nearly elastic spheres,'' Phys. Fluids A {\bf 1}, 2050 
(1989). 
 
\bibitem{Z95}  P. Zamankhan, ``Kinetic theory for multicomponent dense 
mixtures of slightly inelastic spherical particles,'' Phys. Rev. E {\bf 52}, 
4877 (1995). 
 
\bibitem{AW98}  B. Arnarson and J. T. Willits, ``Thermal diffusion in 
binary mixtures of smooth, nearly elastic spheres with and without 
gravity,'' Phys. Fluids {\bf 10}, 1324 (1998). 
 
\bibitem{WA99}  J. T. Willits and B. Arnarson, ``Kinetic theory of a 
binary mixture of nearly elastic disks,'' Phys. Fluids {\bf 11}, 3116 
(1999). 
 
\bibitem{GD299}  V. Garz\'{o} and J. W. Dufty, ``Homogeneous cooling 
state for a granular mixture,'' Phys. Rev. E {\bf 60}, 5706 (1999). 
 
\bibitem{GM84}  S. R. de Groot and P. Mazur, {\em Nonequilibrium 
Thermodynamics} (Dover, New York, 1984). 

\bibitem{BDS97}  J. J. Brey, J. W. Dufty, and A. Santos, ``Dissipative 
dynamics for hard spheres,'' J. Stat. Phys. {\bf 87}, 1051 (1997). 
 
\bibitem{FK72}  J. Ferziger and H. Kaper, {\em Mathematical Theory of 
Transport Processes in Gases} (North-Holland, Amsterdam, 1972). 
 
\bibitem{Zubarev}  D. N. Zubarev, {\em Nonequilibrium Statistical 
Thermodynamics} (Consultants Bureau, NY, 1974). 

\bibitem{BRCG99}  J. J. Brey, M. J. Ruiz-Montero, D. Cubero, and R. 
Garc\'{\i}a-Rojo, ``Self-diffusion in freely evolving granular gases,'' Phys. Fluids {\bf 12}, 876 (2000). 

%\bibitem{MGSD02} J. M. Montanero, V. Garz\'o, A. Santos, and J. W. Dufty,
%``Shear viscosity for a moderately dense granular gas,'' unpublished. 

\bibitem{H00} L. Huilin, L. Wenti, B. Rushan, Y. Lidan, and D. Gidaspow,
``Kinetic theory of fluidized binary granular mixtures with unequal
temperature,'' Physica A {\bf 284}, 265 (2000).

\bibitem{Bird}  G. A. Bird, {\em Molecular Gas Dynamics and the Direct
Simulation Monte Carlo of Gas Flows} (Clarendon, Oxford, 1994). 
\end{references}
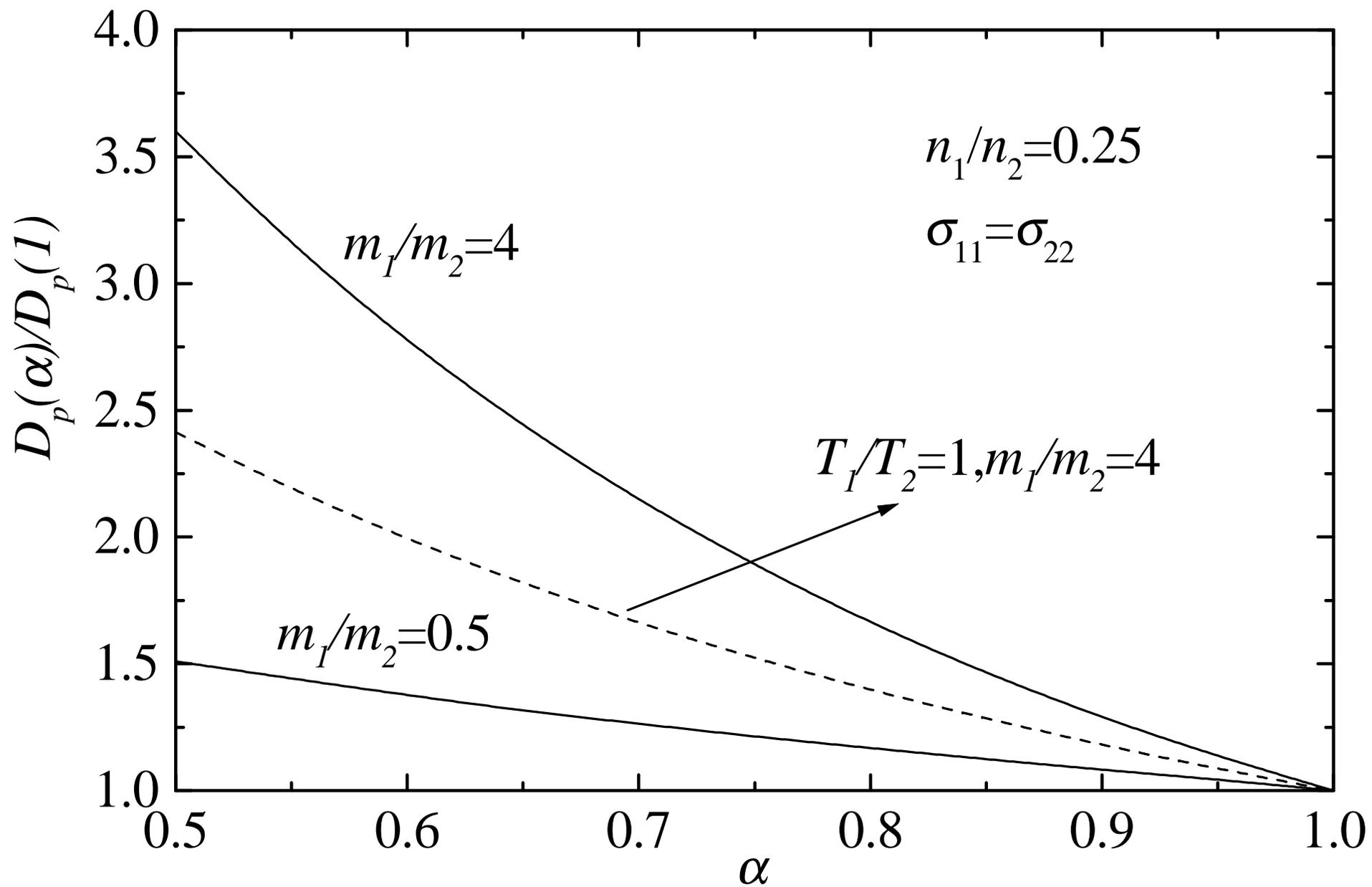
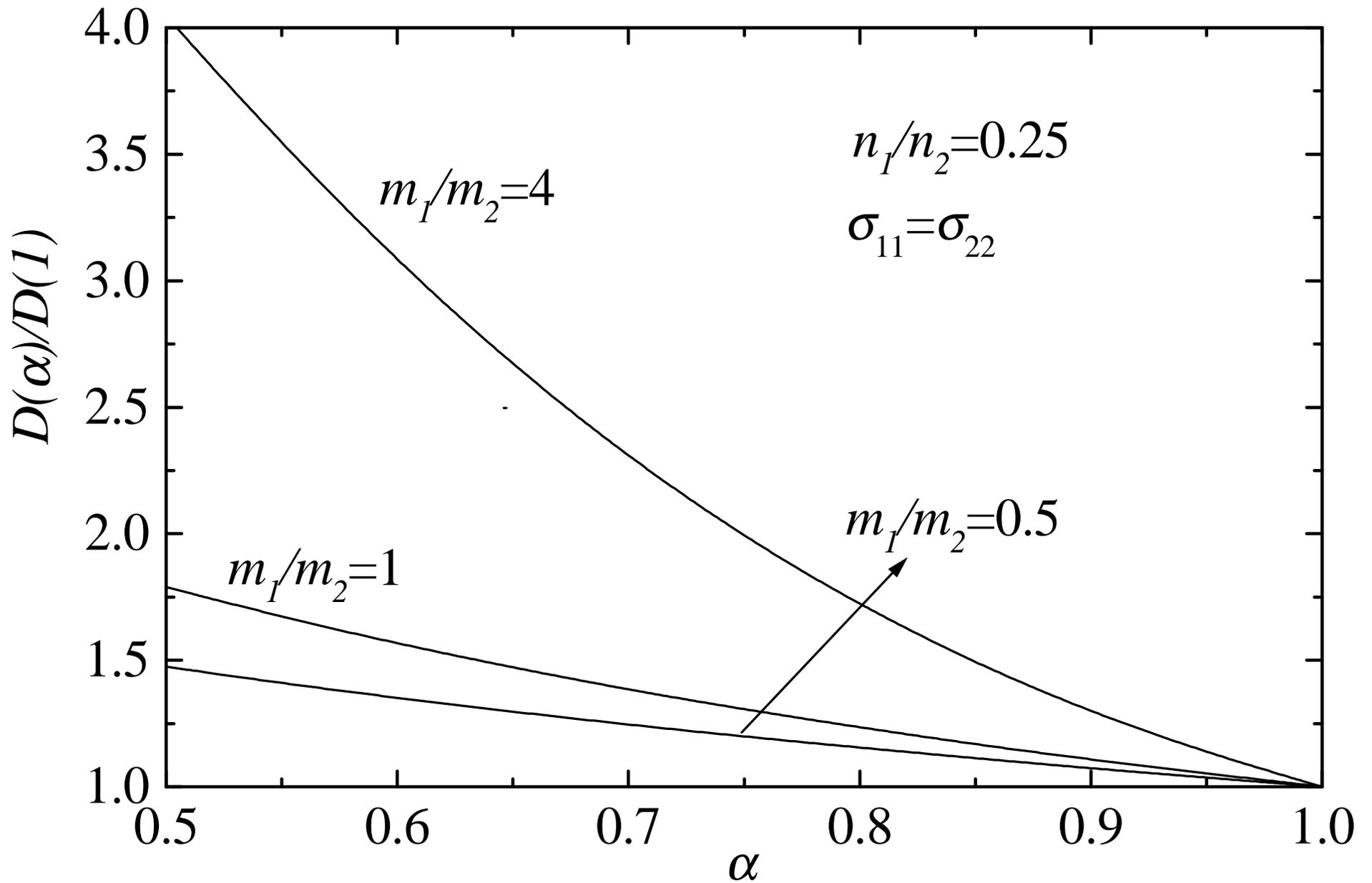
\end{document}